# Electrochemical Performance of Gold Monolayers for Lithium-Ion Batteries: A First-Principles Study

Ajay Kumar[a], Pritam Samanta[b] and Prakash Parida[b]*
[a]Department of Physics, Government Degree College, Chatroo, Kishtwar, 182205, Jammu and Kashmir, India
[b]Department of Physics, Indian Institute of Technology Patna, Bihta, 801106, Bihar, India
*Corresponding authors: pparida@iitp.ac.in

***Abstract***

Being motivated by recent synthesis of a monolayer of gold, named goldene, from the nano-laminated ternary ceramic phase of $Ti_3AuC_2$, we are proposing two phases of goldene viz. goldene-I and goldene-II as anode material for Lithium-Ion batteries using first principles study. This innovative goldene-I monolayer, composed of triangular motifs of gold atoms, exhibits remarkable properties owing to its unique geometric configuration and intrinsic stability. In contrast, a theoretical structure known as goldene-II, featuring a combination of triangular and hexagonal motifs, has been proposed. This structure possesses intrinsic, periodically distributed pores among Au atoms and demonstrates structural integrity and mechanical robustness, even under lithium adsorption. The electronic band spectra and projected density of states reveal the metallic nature of both phases of goldene. Electrochemical evaluations reveal that goldene-II offers favourable lithium-ion adsorption energies, efficient charge transfer, and volumetric capacities. Goldene-I achieves a volumetric capacity of 0.713 Ah/cm³, while goldene-II reaches 0.783 Ah/cm³, confirming its high suitability for lithium storage volumetric capability. Moreover, goldene-I has an ultra-low barrier height of 15 meV, which supports rapid lithium-ion transport.

## 1. Introduction

The global energy industry has traditionally relied on fossil fuels, which release large quantities of greenhouse gases such as carbon dioxide ($CO_2$), carbon monoxide (CO), methane ($CH_4$), and nitrous oxide ($N_2O$), contributing to the greenhouse effect. Additionally, this conventional, non-renewable energy source is finite and is expected to be depleted within the next century due to current overdependence. Therefore, finding alternative, renewable, and eco-friendly energy sources is essential for powering future global industries.[1-3] Thermoelectric materials are one of the green energy harvesting techniques, but their efficiency to convert waste heat into useful energy is very low.[4-9] Therefore, there is an urgent need for energy storage technologies that are both efficient and cost-effective, as many clean energy sources, such as wind and solar, are intermittent and require reliable storage before they can be utilized.

Demand for high-capacity lithium-ion batteries (LIBs) has recently surged, highlighting their features and applications.[10, 11] Initially, battery research was limited, but the growing need for portable electronics and electric vehicles has greatly increased interest. Today, LIBs lead the energy storage market, powering devices from cell phones and computers to electric and hybrid vehicles.[12-15] A significant challenge in LIBs technology is the low theoretical specific capacity of commercially used graphite anodes, limited at 372 mAh/g.[16] Consequently, substantial research efforts are directed toward alternative anode materials having high specific capacity like Si (4200 mAh/g),[17] Sn (994 mAh/g),[18] and $SnO_2$ (782 mAh/g).[19] However, their practical application is hindered by poor cycling stability, resulting from the large volume expansions (100–300% relative to the initial volume) during repeated Li-ion alloying and dealloying, which characterizes their electrochemical behaviour.[20, 21] Mono elemental nanosheets, including graphene,[22, 23] borophene,[24] phosphorene,[25] silicene,[26] germanene,[27] and stanene,[28] have become crucial in nanotechnology due to their exceptional properties. These materials hold significant promise for a wide range of innovative applications across different fields. Recently, Shun Kashiwaya et al. developed a synthetic method to exfoliate goldene from $Ti_3AuC_2$, a nanolaminate ternary ceramic phase. The $M_3AX_2$ family, where M is a transition metal, A belongs to groups 12–16, and X is either carbon (C) or nitrogen (N), has been previously synthesized with silicon (Si) as the A component. In this process, the researchers created a three-dimensional base material by embedding silicon between titanium and carbon layers.[29] At high temperatures, the silicon in titanium silicon carbide ($Ti_3SiC_2$) has been replaced by gold, leading to the formation of titanium gold carbide. Furthermore, their goal is to coat the electrically conductive titanium silicon carbide with gold to enhance contact. By wet-chemically etching the carbide layers using Murakami's reagent with cetrimonium bromide or cysteine, they successfully achieved the exfoliation and left a freestanding single-atom-thick layer of Au.[30] This gold monolayer corresponds to the top layer of the bulk Au(111) face-centred cubic lattice. Li-Ming Yang et al. theoretically explored various 2D structures of gold, including hexagonal close-packed (triangular sheet), square, honeycomb, and tetracoordinate configurations. Among these, only the triangular sheet was found to be stable, while the others exhibited dynamic stability due to the presence of negative frequencies in phonon dispersion calculations.[31] The triangular gold sheet resembles the borophene monolayer, which has been experimentally synthesized using various techniques.[32-34] In addition to the triangular motif, they also reported a meta-honeycomb structure and a mixed triangular-hexagonal motif composed of boron atoms, all of which exhibit metallic properties.

In this study, we investigate the structural, electronic, and electrochemical properties of two types of Au monolayers: goldene-I (a triangular sheet) and goldene-II (comprising triangular and hexagonal motifs). The goldene-II monolayer consists of six Au atoms per unit cell arranged in a rhombus, forming a hexagonal pore ring and a triangular motif. This structure, first theoretically reported by Hui Tang in 2010 for boron, was characterized by a hexagonal hole density ($\eta = 1/6$).[35] They also reported two types of boron sheets, alpha and beta, both composed of triangular and hexagonal motifs. While they are isoelectronic and isomeric, they differ in the arrangement of hexagonal pores, resulting in distinct electronic properties and structural stability.[36, 37]

## 2. Computational Details

We investigate the two phases of gold monolayer using density functional theory (DFT) as implemented in the Vienna Ab initio Simulation Package (VASP).[38] All calculations are performed with the Perdew-Burke-Ernzerhof (PBE) exchange-correlation functional within the generalized gradient approximation (GGA).[39] A vacuum layer of 20 Å is applied along the z-direction to minimise interlayer interactions. The geometries are fully optimized until the atomic forces are below 0.01 eV/Å per atom, with an energy convergence threshold of $10^{-8}$ eV. An energy cutoff of 600 eV is maintained throughout the calculations. For Brillouin zone sampling, we use a 15×15×1 k-point mesh for goldene-I and an 11×11×1 mesh for goldene-II. Phonon dispersion spectra are obtained using the finite displacement method via the phonopy package, employing 4×4×1 and 3×3×1 supercell for goldene-I and goldene-II, respectively.[40] The force constants are calculated through the VASP-phonopy interface.[41] The converged charge density is subsequently used to compute the non-self-consistent electronic band structure and the projected density of states (PDOS). Band structures are plotted along high-symmetry paths using the GGA-PBE functional.[42] We have used LOBSTER programme for calculating integrated crystal orbital Hamiltonian population (ICOHP) calculation for bonding strength analysis.[43, 44] To assess electrochemical properties, we incorporate van der Waals (vdW) interactions between adatoms and the monolayer using the Grimme DFT-D3 method with zero damping. [45] The climbing-image nudged elastic band (CI-NEB) approach is employed to identify the transition state and diffusion barrier for metal adsorption on the goldene monolayers. [46] Thermal stability is evaluated via ab initio molecular dynamics (AIMD) simulations in the NVT ensemble using a 3×3×1 supercell for goldene-I and a 2×2×1 supercell for goldene-II at 500 K, with a time step of 1 fs over a 5 ps duration. [47] The atomic structures and charge densities are visualized using VESTA software, while Xcrysden is utilized to illustrate high-symmetry paths.

## 3. Results and Discussion

### 3.1. Structural stability

In this study, we investigate the experimentally synthesized triangular gold monolayer, referred to as goldene-I, along with the theoretically proposed Au monolayer, goldene-II. [28] **Fig. 1(a)** and **1(b),** depict the top and side views of the optimized geometric structures of both goldene-I and goldene-II monolayers. Both goldene phases adopt a 2D hexagonal Bravais lattice with a rhombic primitive cell. Goldene-I has lattice parameters a = b = 2.73 Å, while goldene-II has a = b = 7.13 Å; the in-plane angle $\gamma$ is 60° in both cases, and a 20 Å vacuum layer is included along the z-direction. Their in-plane crystallographic symmetry is described by the plane group *p6mm*. The structural unit of the triangular phase (goldene-I) consists of a single Au atom, whereas the goldene-II phase contains six Au atoms. The goldene-II phase can be visualized as a structure derived from the triangular phase by removing each seventh Au atom, creating uniform hexagonal pores interspersed with ribbons of triangular motifs.

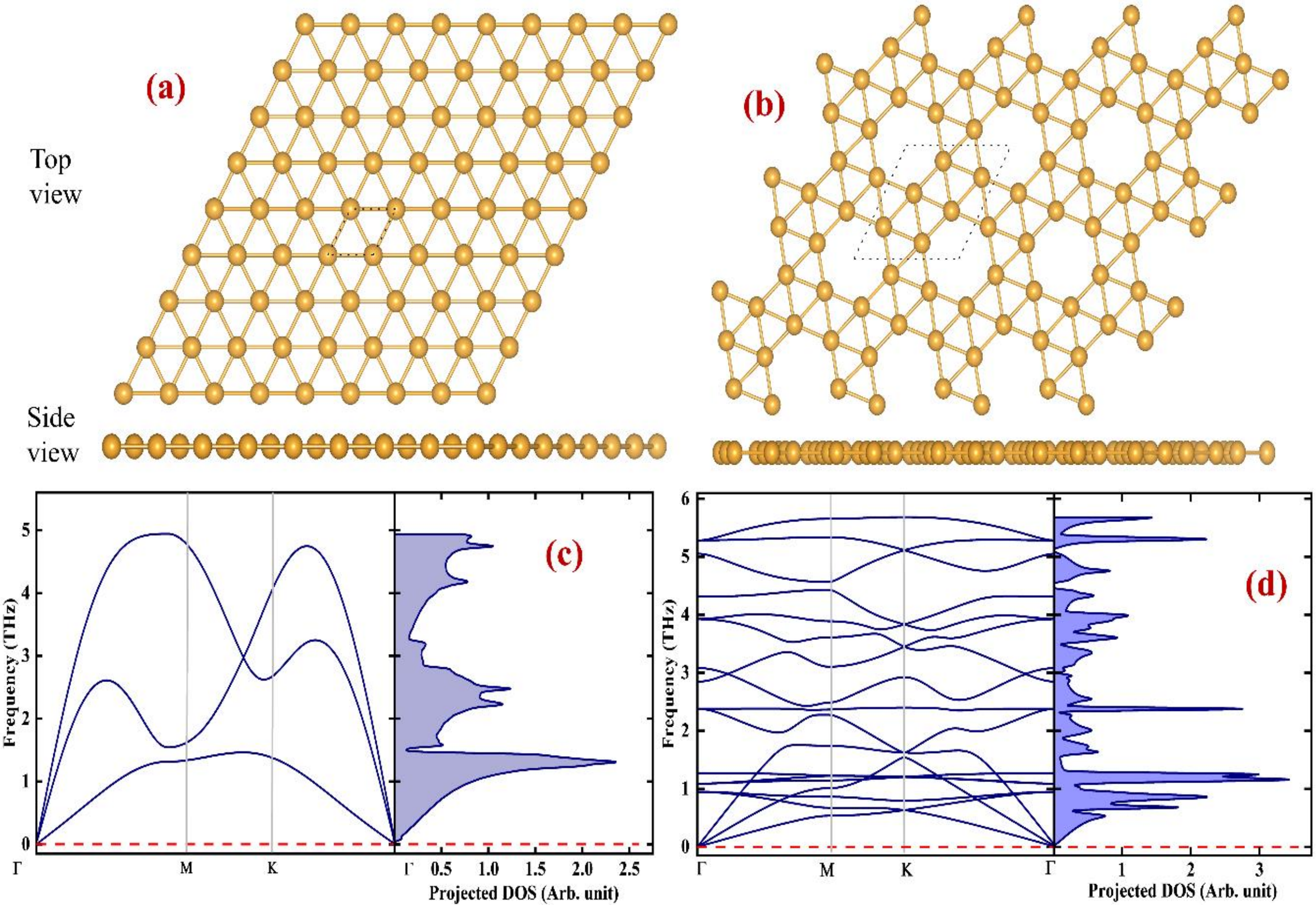

*Fig. 1 (a) and (b) present the top and side views of the fully optimized crystal structures of goldene-I and goldene-II, where dashed lines highlight the rhombus-shaped unit cells. The goldene-I consists of a single Au*

*atom per unit cell, whereas goldene-II comprises six Au atoms per unit cell, reflecting their distinct atomic arrangements. (c) and (d) present the corresponding phonon band dispersions along high-symmetry directions of the Brillouin zone together with the phonon density of states for the goldene-I and goldene-II monolayers, providing insight into their dynamical stability.*

To evaluate the structural stability, we calculated the formation energy $E_{form}$, defined as

$$\mathrm{E_{form}} = (\mathrm{E_{monolayer}} - \mathrm{m}\mu_{\mathrm{Au}})/\mathrm{m}, \tag{1}$$

where $\mathrm{E_{monolayer}}$ and $\mu_{\mathrm{Au}}$ represent the total energy of a unit cell of the goldene-I/goldene-II and the chemical potential of Au extracted from a stable bulk face-centred cubic (FCC) phase of gold, [48] respectively. And 'm' denotes the number of Au atoms in the unit cell. The formation energies of the goldene-I and goldene-II monolayers are -2.43 eV and -2.11 eV per atom, respectively, consistent with previous studies for the goldene-I phase. [31] Further, the uniform hexacoordinate lattice in goldene-I monolayer, characterized by a closely packed structure, maximizes in-plane chemical bonding, enhancing its stability. However, the hexagonal motifs in goldene-II slightly destabilize the monolayer due to the differing coordination of Au atoms in comparison to goldene-I. Both monolayers exhibit phonon modes with positive frequencies across the Brillouin zone (BZ), indicating their dynamical stability. The goldene-I monolayer, with a single atom per unit cell, shows three acoustic modes. In contrast, the goldene-II monolayer, which contains six atoms per unit cell, exhibits three acoustic modes and 15 optical modes. As illustrated in **Fig. 1(c)** and **Fig. 1(d)**, the phonon dispersion curves confirm the stability of these two monolayers. Further, the goldene-II monolayer exhibits hybridization between the optical and acoustic branches, along with a flat dispersion of phonon states. This leads to a sharp peak in the phonon density of states within the 1-1.5 THz frequency range. We have calculated the phonon band centre (average vibrational frequency), $\omega_{av} = \frac{\int \omega \times g(\omega) d\omega}{\int g(\omega) d\omega}$ (here $g(\omega)$ is the density of states of phonon) to get insight about diffusion of Li in anode materials. [49] A lower phonon band centre in anode materials generally promotes faster Li diffusion because of softer lattice. A softer lattice generally has a lower migration energy barrier for Li ion. In our case, we found $\omega_{av}$ = 2.25 THz for goldene-I and $\omega_{av}$ = 2.5 THz for goldene-II. As goldene-I is softer compared to goldene-II, it may be expected that Li ion will face a lower migration barrier height in goldene-I during diffusion process. This qualitative behaviour will be further confirmed by calculating the barrier height using CI-NEB approach in the section **3.3.5**. Another important aspect is that strong scattering arising from acoustic–optical phonon hybridization significantly suppresses heat transport. In 2D MXenes, such as $Ti_3C_2T_x$ (here T denotes surface terminal groups viz. -O, -OH, -F) extensive mode mixing and anharmonicity result in an ultralow lattice thermal conductivity (2-5 $Wm^{-1}K^{-1}$), which helps prevent local overheating, reduces thermal runaway risks, and improves the long-term stability of LIB electrodes.[50]. A similar behaviour is observed in goldene-II, where the pronounced overlap between acoustic and optical phonon branches may suggest reduced lattice thermal conductivity. This implies that goldene-II may be more effective in mitigating thermal runaway and overheating issues during battery operation. The mechanical properties of 2D materials, including Young's modulus (Y) and Poisson's ratio (ν), are crucial to verifying mechanical strength. To evaluate these properties, we first calculate the elastic constants ($C_{ijkl}$) using the strain-energy method, which follows generalized Hooke's law, expressed as:

$$\sigma_{ij} = C_{ijkl}\epsilon_{kl} \; ; \; C_{ijkl} = \frac{1}{2}\frac{\partial^2 U}{\partial \epsilon_{ij} \partial \epsilon_{kl}} \; (where \; i, j, k, l = 1,2,3), \tag{2}$$

where σ is second-rank stress and $\epsilon$ represents the strain tensors, and $C$ is the fourth-rank elastic coefficients. These elastic constants are nomenclatured as $C_{ijkl} \rightarrow$ $C_{pq}(p,q=1,2,3,4,5,6)$ in Voigt notation[51, 52]. For illustration *ij→p* is given *as 11→1, 22→2, 33→3, 21=12→4, 13=31→5, 23=32→6.*

The elastic strain energy per unit area (U) can be defined as;

$$U(\epsilon_{11},\epsilon_{22}) = \frac{1}{2}C_{11}\epsilon_{11}^2 + C_{22}\epsilon_{22}^2 + C_{12}\epsilon_1\epsilon_2 + \cdots \quad (3)$$

Here, $\epsilon_{pq}$ represents the strain along the x and y-axes, and $C_{11}, C_{22}, C_{12}$ *and* $C_{66}$ are elastic constants in Voigt notation. Based on the obtained elastic constants, we then determine Y and ν using the relationships defined in **eq. (4)**, which are specific to the hexagonal symmetry of the 2D system. [4, 52]

$$\mathrm{Y} = \frac{C_{11}^2 - C_{12}^2}{C_{11}} \;\; \& \;\; \nu = \frac{C_{12}}{C_{11}}. \quad (4)$$

**Fig. 1(a),** illustrates the top and side views of the goldene-I and goldene-II monolayers. Both monolayers possess a hexagonal lattice, which necessitates the calculation of three elastic constants: $C_{11}$, $C_{12}$, and $C_{66}$. The elastic constants were obtained using the strain–energy method. For the goldene-I monolayer, the calculated in-plane elastic constants $C_{11}$, $C_{12}$, and $C_{66}$ are 115.86, 46.84 and 34.50 N/m, respectively. In contrast, the goldene-II monolayer exhibits values of $C_{11} = 86.57$ N/m, $C_{12} = 47.23$ N/m, and $C_{66} = 9.67$ N/m. Furthermore, for both monolayers, the conditions $C_{11}$, $C_{12}$, $C_{66} > 0$ and $C_{11} > C_{12}$ are satisfied, confirming their mechanical stability. [6, 53]

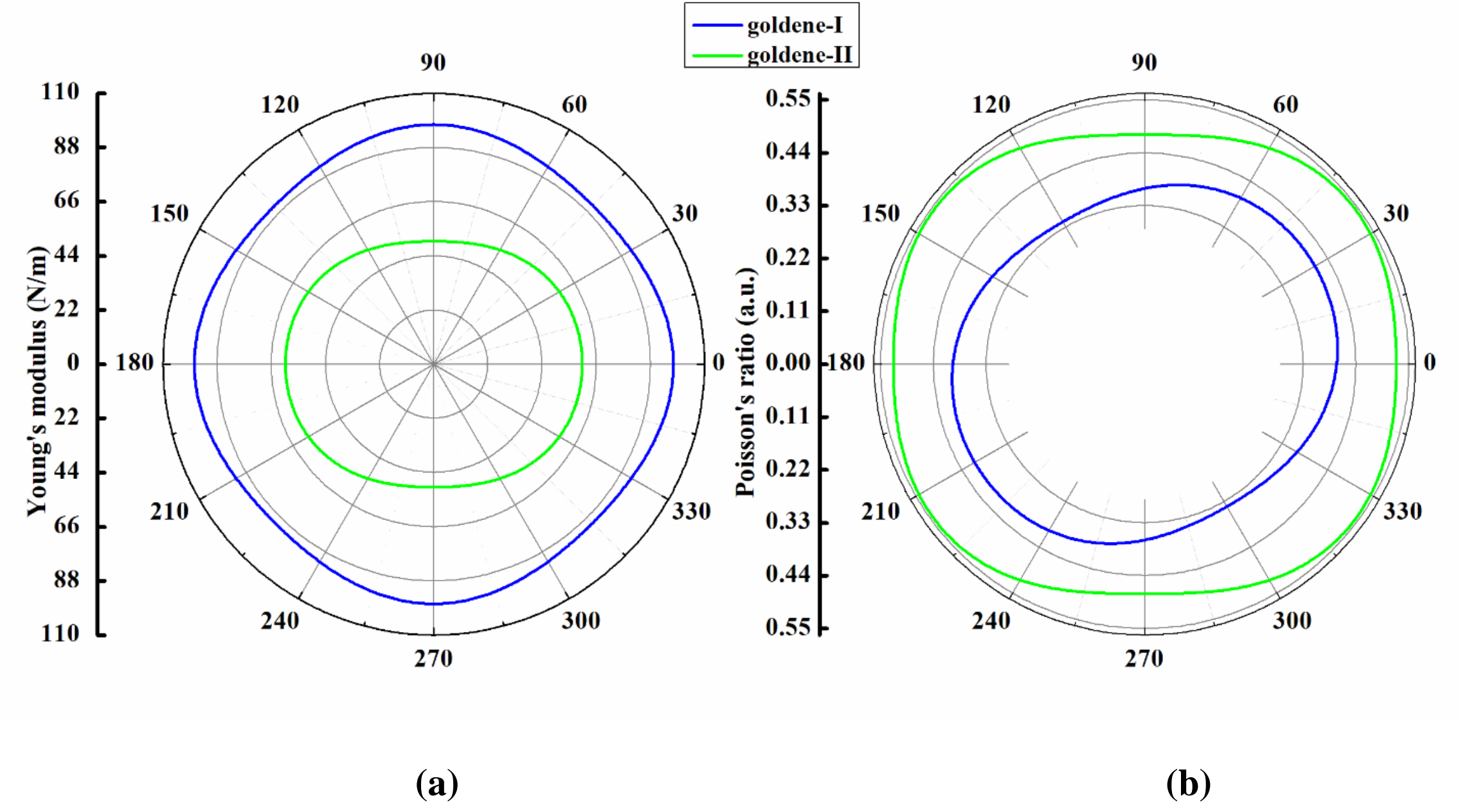

**(a)** **(b)**

***Fig 2:*** *Angle-dependent (a) Young's modulus (Y) and (b) Poisson's ratio (ν) of goldene-I and goldene-II monolayers.*

Furthermore, Young's modulus and Poisson's ratio have been calculated using the elastic constants in **eq. 2, 3 and 4**. As shown in **Fig. 2**, the maximum value of Young's modulus for goldene-I and goldene-II is 96.92 N/m and 60.80 N/m, respectively. Conversely, an increasing trend is noted for Poisson's ratio, which is 0.40 for goldene-I and 0.54 for goldene-II. It is well known that the incompressibility of a material is linked to its Poisson ratio, with values reaching at least 0.5, indicating that the material can be considered incompressible. In our studied systems, the Poisson's ratios for both monolayers are 0.40 and 0.54, suggesting that both monolayers exhibit low ductility and flexibility.

Our theoretical calculations indicate that both phases of the gold monolayer exhibit comparable stability. Since goldene-I has already been experimentally synthesized, it is essential to assess the experimental feasibility of goldene-II. This feasibility is strongly supported by its close structural analogy to $\chi_6$ borophene, a hollow-hexagon, η = 1/6 vacancy-ordered polymorph that has been successfully synthesized on an Ir (111) substrate.[54] Goldene-II adopts the same vacancy-engineering principle, where removal of every sixth atom from the triangular goldene-I lattice generates an ordered hollow-hexagon network. Considering that freestanding-like goldene monolayers have been experimentally achieved and that Au exhibits even richer 2D polymorphism than boron, similar substrate-templated growth and vacancy ordering should be transferable to gold. Furthermore, our calculations confirm that goldene-II meets all the key stability criteria, like energetic, dynamical, mechanical, and thermal, further reinforcing its feasibility. Collectively, these factors indicate that goldene-II is a realistic candidate for synthesis under ultra-high vacuum conditions on substrates such as Ir (111), Cu (111), or Ag (111).

### 3.2. Electronic properties

We next investigated the electronic properties of these two goldene monolayers. For this purpose, we simulated the band structures of the monolayers as well as their density of states (DOS). As shown in **Fig. 3 (a),** two bands cross the Fermi level ($E_F$), demonstrating the metallic character of goldene-I. In detail, $d_{x^2-y^2}$ orbitals make the main contribution around the $E_F$ as depicted in the projected density of states (PDOS) plot given in the right panel of **Fig. 3**. On the other hand, the goldene-II conduction band make a hole pocket around the **M** high symmetry point. The valence band touches the Fermi energy at the **Γ** point, indicating a semi metallic nature. Projected DOS suggest that the $d_{x^2-y^2}$ and $d_{xy}$ have a major contribution to the valence band. In contrast, the conduction band is hybridized of $d_{x^2-y^2}$ and $d_{z^2}$ orbitals.

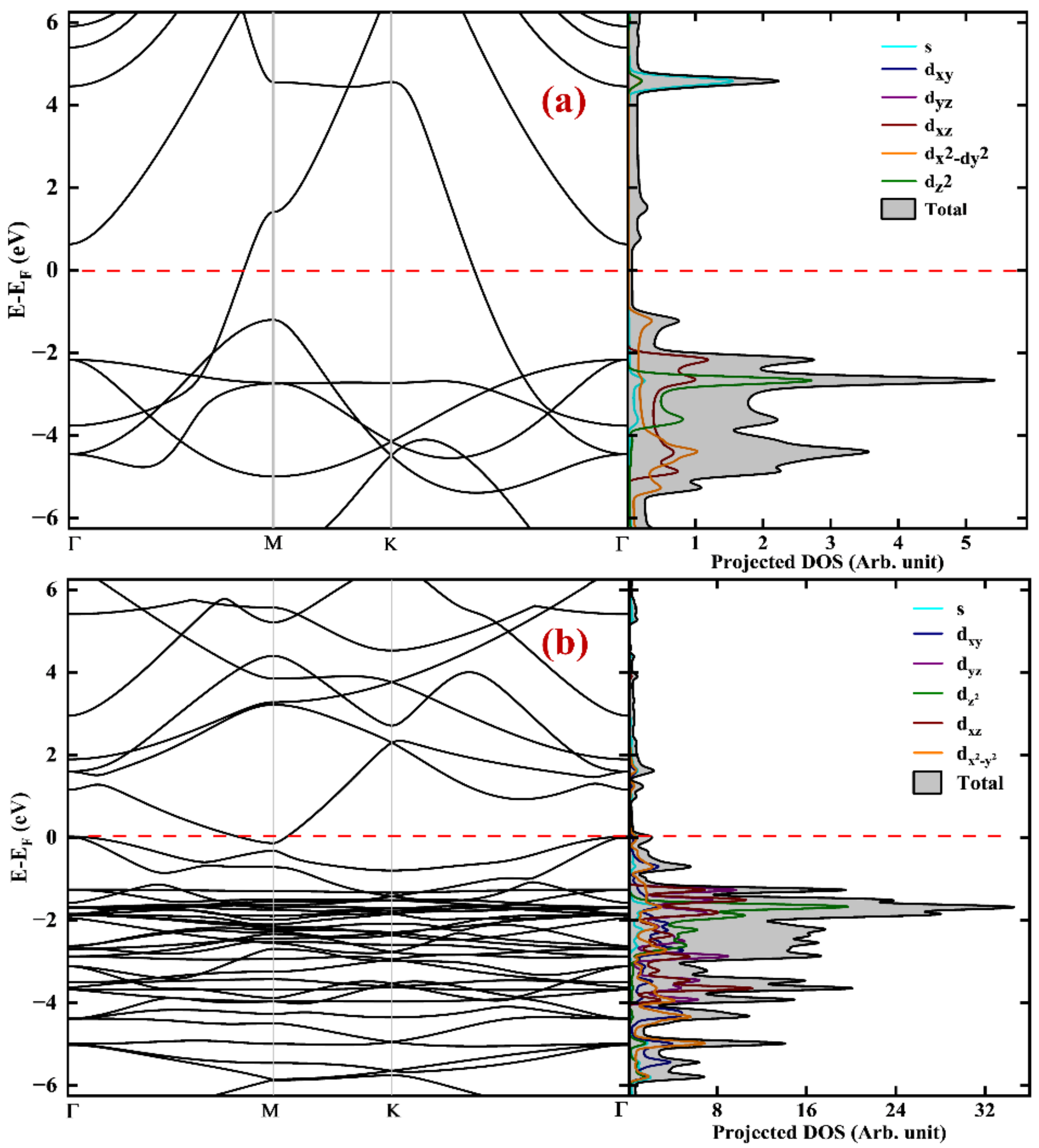

*Fig. 3 Electronic bands spectra of (a) goldene-I and (b) goldene-II along with their projected density of states.*

The goldene-I, being metallic, offers higher intrinsic electronic and ionic conductivity that facilitates fast charge transfer; however, its high free-electron density may promote stable solid electrolyte interphase (SEI) instability and structural perturbations during cycling. In contrast, goldene-II exhibits a semi-metallic nature with high carrier mobility and band edges near the Fermi level, ensuring adequate electrical conductivity while suppressing excessive electron density. The commercial used anode material graphite is also semi-metal in nature.[16] This semi-metallic framework favors uniform Li adsorption, mitigates dendrite formation and volume-change-induced degradation, stabilizes the SEI, and enables a higher reversible Li storage capacity, thereby making goldene-II a more reliable and durable anode material for long-term electrochemical applications.

### 3.3. Electrochemical properties

#### 3.3.1. Adsorption energy

To explore the potential of goldene-I and goldene-II monolayers as anode materials for LIB rechargeable batteries, we must first investigate the adsorption and diffusion of Li atoms on the anode surface, which facilitates ion transport. For this, a single Li atom has been adsorbed at different sites on a 3×3×1 supercell of goldene-I and a 2×2×1 supercell of goldene-II. Based on crystal symmetry, goldene-I has three potential adsorption sites, as illustrated in **Fig. 4(a)**. The lithium atom positioned directly above Au atoms are designated as Au-sites, those at the centre of triangular motifs are called T-sites, and a third type, located over the Au-Au bond, is termed the B-site i.e. bridge site of two Au atoms. By introducing ordered pores into the goldene-I, we propose a new phase of goldene, called goldene-II. It has five possible symmetric adsorption sites for adsorption of Li as shown in right panel of **Fig. 4(b).** Among these five possible adsorption sites, three sites (Au-site, T-site, and B-site) are already present in goldene-I. Additionally, goldene-II features extra two adsorption sites at the centre of the hexagonal motif (H-site), and at the bridge site (B'-site). Note that, B'-site is not regular B-site present in goldene-I. This B'-site near the hexagon motif(H-site) is created on introduction of ordered pores into regular goldene-I phase. Furthermore, the adsorption energy has been calculated by the following equation[55];

$$\mathrm{E_{ads}} = \mathrm{E_{Li@goldene}} - \mathrm{E_{goldene}} - nE_{\mathrm{Li},} \tag{5}$$

where $\mathrm{E_{Li@goldene}}$ is the total energy of Li-adsorbed goldene, $\mathrm{E_{goldene}}$ is the energy of the pristine goldene monolayer and $\mathrm{E_{Li}}$ is energy of the isolated Li atom. [55, 56] 'n' is the number of lithium atoms adsorbed on the monolayer. The calculated adsorption energy by **eq. 5,** shows the T-, Au- and B-sites for goldene-I are favourable because of negative adsorption energy. We have plotted histogram for adsorption energy of lithium atom at adsorption sites in the right panel of **Fig. 4,** for both goldene-I and goldene-II. For goldene-I, out of these three sites, the T site is the most stable site for Li adsorption, followed by B-site, and the least favourable is the Au-site, as shown in **Fig. 4(a).**

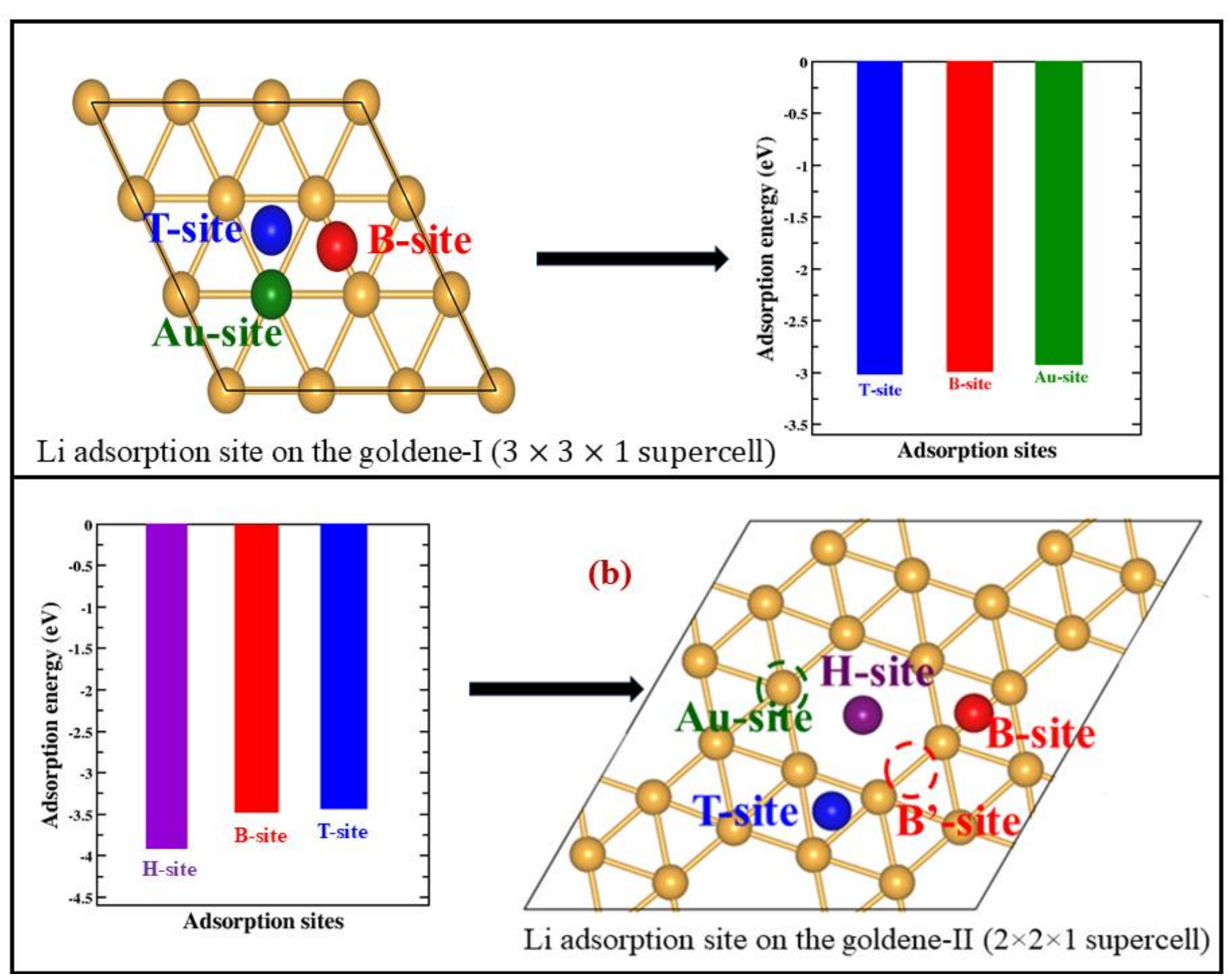

***Fig. 4*** *Different adsorption sites and their corresponding energy of (a) goldene-I and (b) goldene-II.*

In contrast, we have found there are three favourable adsorption sites for goldene-II phase viz, H-, T-site, and B-site after geometry optimization. Au-site and B'-site are not stable adsorption sites. After geometry optimization Li atom is displaced from Au-site to B-site, and similarly Li atom is displaced from B'-site to T-site. Notably, among three stable adsorption sites the H-site is the most favorable adsorption site, exhibiting the strongest binding affinity for the Li atom. Following the H-site, the B-site is the second most stable site and T-site is the lowest stable adsorption site having negligibly higher adsorption energy than B-site. The goldene-II has preferential adsorption on the H-site, strongly supporting the literature that the existence of suitable pores would enhance the Li adsorption.[37] **Fig. 4(a)** and **4(b)** show the colour code of each site along with their corresponding adsorption energy histogram for different sites for both the phases of goldene. For goldene-I, the preferred adsorption sites are the T-site, followed by the B-site for Li atom adsorption. Au-site is the least stable site in goldene-I phase. In goldene-II, Li atom adsorption is most favourable at the H-site, followed by the B-site and the T-site. The adsorption energy at each site, along with their corresponding vertical height (the distance between the adsorbed Li atom and the monolayer), is detailed in **Table 1**.

***Table 1:*** *The most favourable adsorption sites and their corresponding adsorption energies, vertical distance between the Li atom and monolayer (h) and charge transfer for goldene-I and goldene-II monolayers.*

| Phases | Sites | $E_{ads}$ (eV) | h (Å) | Charge transfer ($e^-$) |
|---|---|---|---|---|
| goldene-I | T-site | -3.036 | 1.96 | 0.89 |
| | B-site | -3.021 | 2.11 | 0.88 |
| | Au-site | -2.856 | 2.28 | 0.906 |
| goldene-II | H-site | -4.06 | 1.24 | 0.96 |
| | B-site | -3.47 | 2.33 | 0.89 |
| | T-site | -3.42 | 1.95 | 0.88 |

We found adsorption energies for T-site, B-site, and Au-site as −3.036 eV, −3.021 eV, and −2.85 eV for goldene-I. For goldene-II, these are −4.06 eV, −3.47 eV, and −3.42 eV for H-site, B-site, and T-site, respectively. We note that, the above adsorption energies were calculated using energy of an isolated Li atom as the reference energy. We have also calculated the adsorption energies using energy reference state of bcc Li. The calculated adsorption energies using energy reference state of bcc Li (chemical potential of lithium, $\mu_{Li} = -1.92\ eV$) shows that the adsorption is moderate on the goldene monolayers. When we use energy of bcc Li, for goldene-I, the calculated adsorption energies at the T-site, B-site, and Au-site for goldene-I are −1.14 eV, −1.12 eV, and −0.964 eV, respectively. Similarly, for goldene-II, we have found the adsorption energies at the H-site, B-site, and the T-site are −2.16 eV, −1.57 eV, and −1.52 eV, respectively. These adsorption energies are relatively stronger than the bulk gold which is inert toward Li. It is known that bulk gold is naturally considered as a noble metal with weak chemical affinity toward Li. But as we decrease the dimensionality of the system, the interaction toward lithium changes significantly. This strong binding of lithium originates in the goldene-I and goldene-II monolayers from the modified electronic structure and enhanced surface reactivity of the goldene monolayers due to the reduced dimensionality of the goldene monolayer. In goldene monolayer, the lithium atoms bind at multiple neighbouring gold atoms which gives rise to such moderate binding of lithium atoms. From the PDOS plot in **Fig. S1** (see **Fig. S1** in **SI**), it is observed that the delocalized Au 6s and partially hybridized 5d states near the Fermi level provide efficient electronic screening and allow the goldene monolayer to readily accommodate electrons donated by Li atoms. PDOS plot in **Fig. S1** indicates significant hybridization between Li (2s) and Au (6s, 5d) orbitals near the Fermi level. From the Bader charge transfer analysis plotted in **Fig. 5**, it is confirmed that there is a substantial charge transfer from Li to the goldene monolayers of an amount of $0.89\boldsymbol{e}^-$ and $0.96\boldsymbol{e}^-$ for goldene-I and goldene-II monolayer, respectively. This substantial Bader charge transfer indicates a pronounced ionic character which stabilizes the lithium atoms on the goldene surface resulting in such a strong adsorption compared to the inert bulk fcc gold. Martins et al. have reported similar strong adsorption energy trend while lithium sulfide and polysulphide clusters ($Li_2S$, $Li_2S_2$, $Li_2S_4$, $Li_2S_6$, $Li_2S_8$, and $S_8$) bind to goldene-I. [60] They have examined adsorption behaviour of lithium sulfide and polysulfide clusters ($Li_2S$, $Li_2S_2$, $Li_2S_4$, $Li_2S_6$, $Li_2S_8$, and $S_8$) on goldene-I surface for understanding the anchoring properties of goldene-I monolayer. They have shown that goldene-I binds strongly all the lithium sulfides except the polysulfide ($S_8$) with an adsorption energy ranging from −4.29 eV to −1.90 eV. These strong binding induces a substantial charge transfer of $0.92\boldsymbol{e}^{-}$ from Li-S clusters to goldene substrate. This charge transfer is almost similar to our study where Li atom transfers a maximum of $0.89\boldsymbol{e}^{-}$ to the goldene-I at the most stable T-site and $0.96\boldsymbol{e}^{-}$ to the goldene-II monolayer at the most stable H-site. They showed that, Au d states are located at the Fermi level for goldene-I, which is the same in our case. The above reported study of anchoring of Li-S clusters on goldene substrate also shows that goldene monolayer is a chemically active substrate, which further supports our obtained strong adsorption energy for Li on both goldene monolayers.

### 3.3.2. Electron density difference (EDD)

To facilitate the redox reaction at the anode, charge transfer from the Li atom to the substrate plays a crucial role. To investigate this, we calculated the electron density difference using **eq. 5**. The EDD has been calculated as the difference of electron density between Li-adsorbed goldene monolayer, and pristine goldene and Li atom. Mathematically, it is defined as the following formula;

$$\rho_{(net)} = \rho_{(Li@goldene)} - (\rho_{goldene} + \rho_{Li}), \quad (6)$$

where $\rho_{(Li@goldene)}$ and $\rho_{goldene}$ are the electron densities of the optimized Li-adsorbed goldene monolayer and pristine goldene monolayer, respectively, and $\rho_{Li}$ is the electron density of the Li atom.

**(a)**

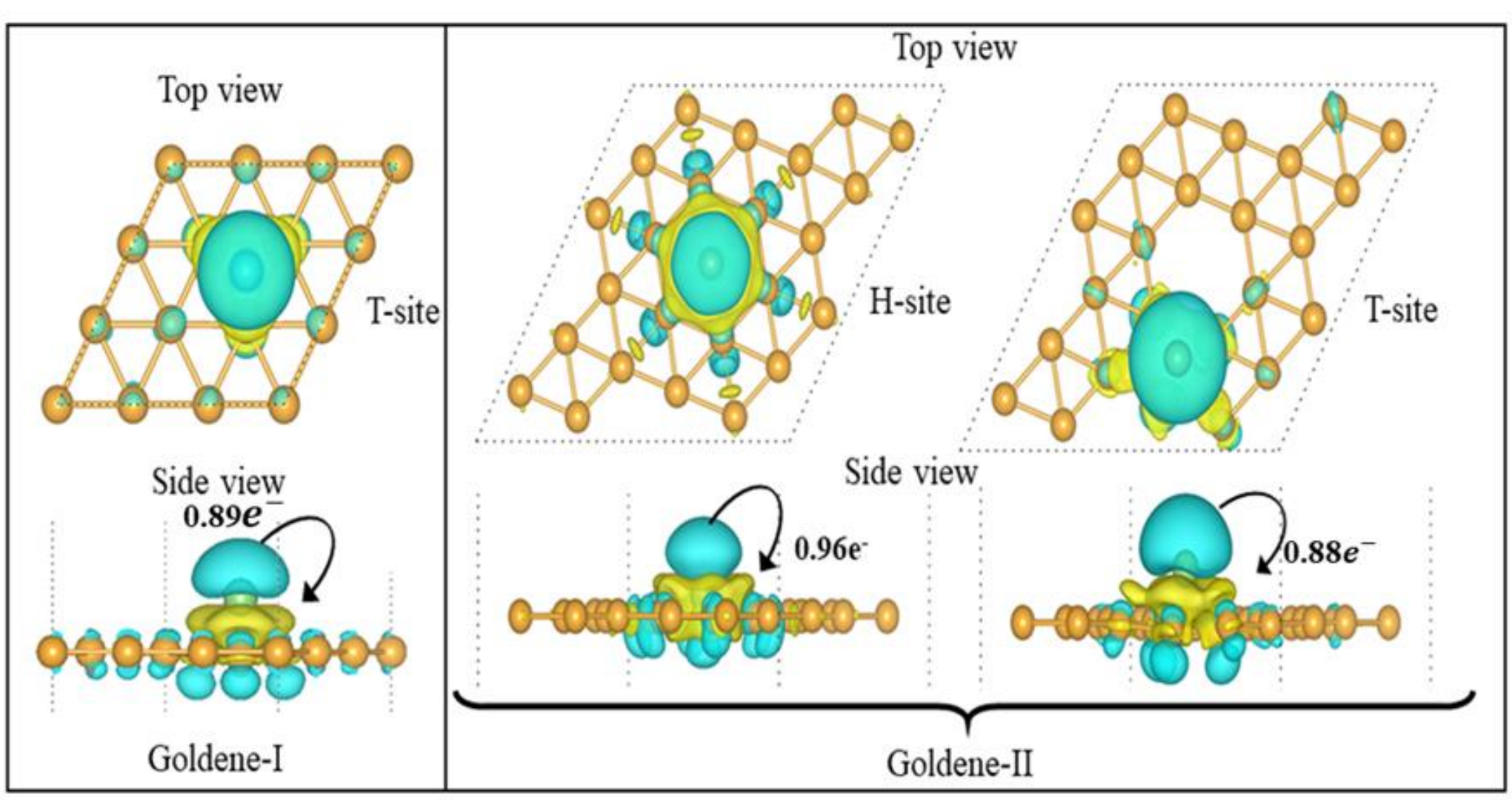

**(b)**

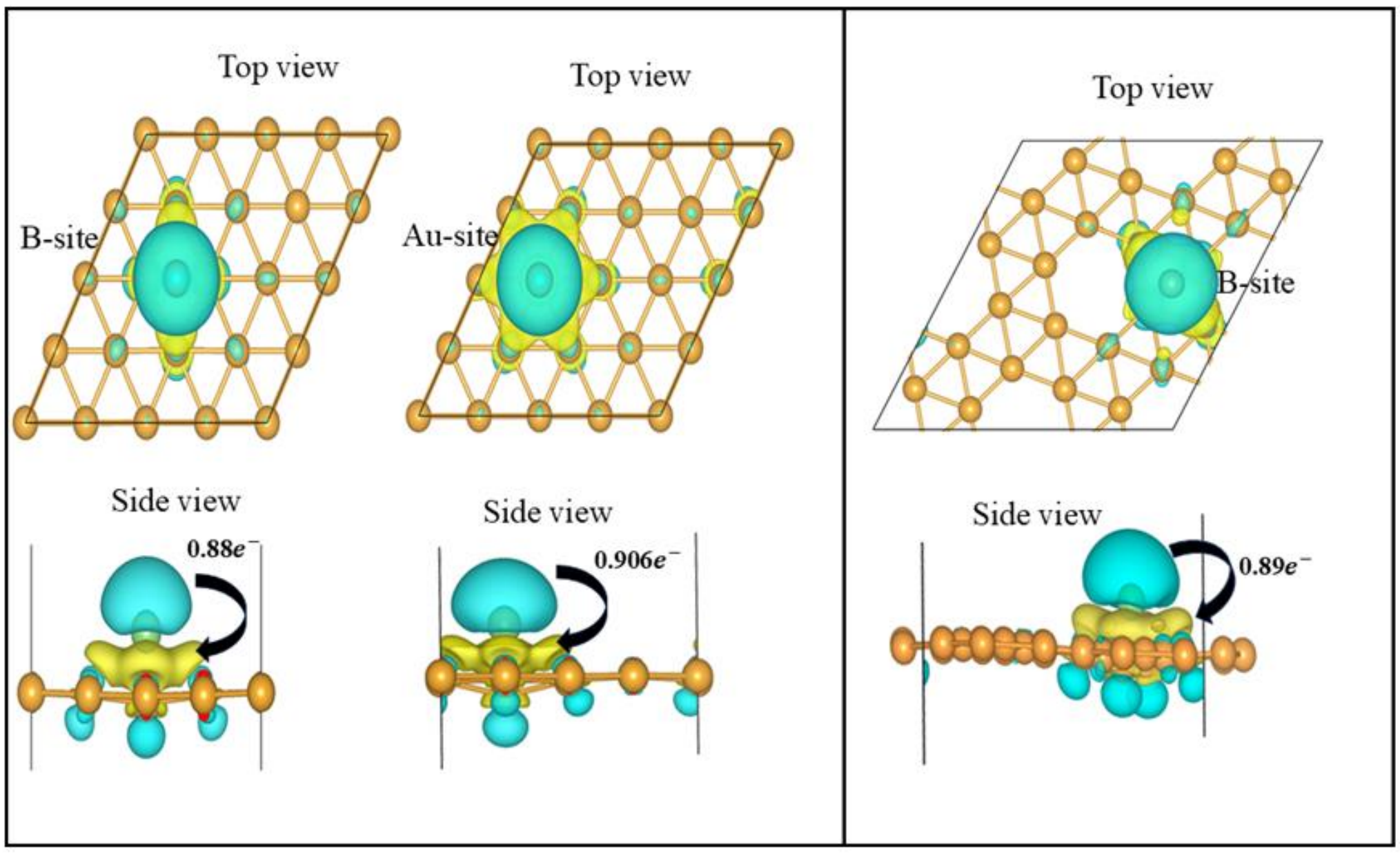

***Fig. 5*** *(a) Electron density difference plots (EDD) for T-site of goldene-I (left panel) and H-site and T-site of goldene-II (right panel) and (b) electron density difference plots for B-site and Au-site of goldene-I (left panel) and B-site of (right panel) goldene-II. Here upper portion and lower portion of each of the panel describes the top view and side view EDD plots with an iso-surface value of 0.002* $e\text{Å}^{-3}$*. The cyan colour and yellow colour indicate charge depletion and charge accumulation region respectively.*

We have shown the schematics of the EDD plots for all the adsorption sites for both goldene-I and goldene-II phases in **Fig. 5**. In these plots, cyan regions denote areas of electron depletion, while yellow regions signify electron accumulation. The cyan envelope around the Li atom indicates a positive charge, suggesting that the Li atom donates electrons to the substrate in both goldene-I and goldene-II phases. In the H-site of goldene-II, the electron distribution is symmetrical over all the Au atoms. The EDD plot for the T-site in goldene-I displays symmetry, attributed to the symmetric atomic geometry of its triangular motifs. However, in goldene-II, the electron distribution at the T-site is asymmetric due to the presence of a hexagonal pore, which distorts the symmetry of the triangular motifs. To quantify the charge transfer, we employed Bader charge analysis. [57] The results reveal a noticeable charge transfer from the Li atom to the monolayers. For goldene-I, the charge transfer is approximately $0.89\boldsymbol{e}^{-}$ at the T-site. In the case of goldene-II, the corresponding values are $0.88\boldsymbol{e}^{-}$ at the T-site and $0.96\boldsymbol{e}^{-}$ at the H-site. The substantial electron transfer at both the T-site and H-site suggests strong ionic interactions between the adsorbed Li atoms and the goldene substrate. **Table 1,** summarizes the adsorption energy, vertical height, and charge transfer for different sites in both the goldene phases.

### 3.3.3. Multilayer Lithium Adsorption on goldene

Storage capacity is a key indicator of the power density in metal-ion battery electrodes and is strongly influenced by the concentration of metal ions adsorbed onto the electrode material. To assess the embedding mechanism and theoretical capacity of Li atoms on goldene, the average adsorption energy ($\mathrm{E}_{ave}$) has been calculated by progressively adding multilayers of Li atoms to both the upper and lower surfaces of a goldene monolayer. The average adsorption energy ($E_{ave}$) is calculated using the following equation:[55, 58]

$$E_{ave} = \frac{E_{Li_x goldene} - E_{goldene} - x\, E_{Li}}{x}. \quad (7)$$

Here, $E_{Li_x goldene}$ , $E_{goldene}$ and $E_{Li}$ denote the total energies of the Li-adsorbed goldene monolayer, pristine goldene monolayer, and energy of isolated Li atom, respectively, while '*x*' represents the number of adsorbed Li atoms. Here, a large negative $E_{ave}$ indicates that additional Li atoms can still be adsorbed, while smaller negative or positive values suggest that further Li insertion is energetically unfavourable. To explore this, 3×3×1 and 2×2×1 supercells of goldene-I and goldene-II, respectively, were used to evaluate storage capacity.

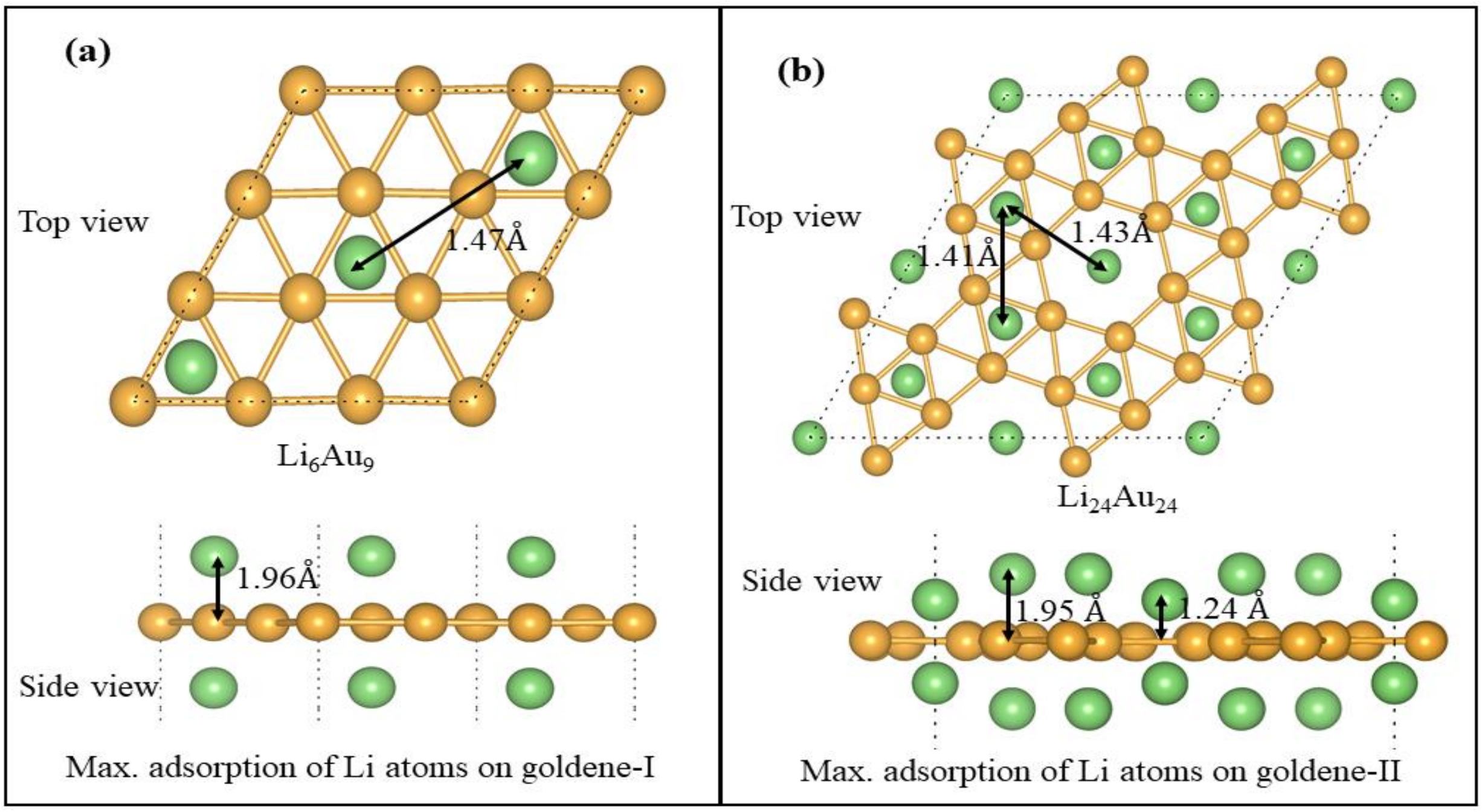

***Fig. 6*** *The top and side views of optimized structures of maximum Li adsorbed on (a) goldene-I and (b) goldene-II monolayers.*

We have calculated adsorption energies for different concentration of lithium on both the goldene-I and goldene-II monolayer phases. In **Fig.7,** we have plotted adsorption energy versus concentration of lithium atoms. In the **Fig. 7,** ‘x’ represents the number of lithium atoms stored on both the phases of goldene. We have stored lithium atoms for diferent number of x, where it ranges from x=1 to x=6 for goldene-I. For all the concentrations, the adsorption energy is negative indicating that lithium adsorption on goldene-I monolayer up-to two layers is energetically favourable. For goldene-II, x takes the values from x=1 to x=24. We note that, a maximum of 6 lithium atoms can be adsorbed on goldene-I monolayer; while maximum 24 numbers of lithium atoms can be adsorbed on the goldene-II monolayer. We see the adsorption energy is negative for all the concentration of lithium atoms. From the right panel of **Fig. 7,** we find the adsorption energy for different number of lithium atoms up-to four layers are negative, and also less than −1.92 eV/atom (cohesive energy of Li atom in its stable bcc phase). [59] The adsorption energy for all the concentrations up-to two layers for goldene-I and four layers for goldene-II are negative. Since the adsorption energy for different number of Li atoms are less than the cohesive energy of metal Li atom, the lithium adsorption on the goldene monolayer matrix is energetically favoured. [59] This energetically favoured lithium adsorption hinders Li metal clustering on the monolayer i.e. reduction of dendrite growth on the matrix. So, the adsorption energy plot indicates that lithium atom stored on goldene monolayer will provide good storage capacity thus increasing the electrochemical performance of goldene monolayer as anode material. For goldene-I, the most favourable adsorption site was identified as the T-site after comparing adsorption energies at various positions. This configuration supports only two layers of Li atoms: one layer of 3 Li atoms on the top surface and another layer of 3 Li atoms on the bottom surface per supercell. Attempting to occupy the third most favourable site (Au-side) results in a very small Li-Li distance, causing the average adsorption energy tend to zero. Consequently, goldene-I is limited to adsorbing just two Li layers, as that happened in graphene. The double-layer Li adsorption configuration for goldene-I, depicted in **Fig. 6(a),** yields negative $E_{ave}$ values. In contrast, goldene-II offers both H-sites and T-sites for adsorption, featuring hexagonal pores of varying vertical heights that facilitate multilayer adsorption. The adsorption process begins by filling the H-site with one layer on either side of the goldene monolayer, at a vertical height of 1.24 Å. Subsequently, the T-sites, located 1.94 Å from the monolayer, accommodate symmetrical third and fourth layers on both sides, as can be seen in **Fig. 6(b)**. The calculated $E_{ave}$ remains negative across all the four layers, though its magnitude decreases progressively with each additional layer, indicating that adsorption up to four layers remains energetically favourable. These results highlight goldene-II is superior to support multilayer Li adsorption compared to goldene-I, enhancing its potential for high-capacity electrode applications.

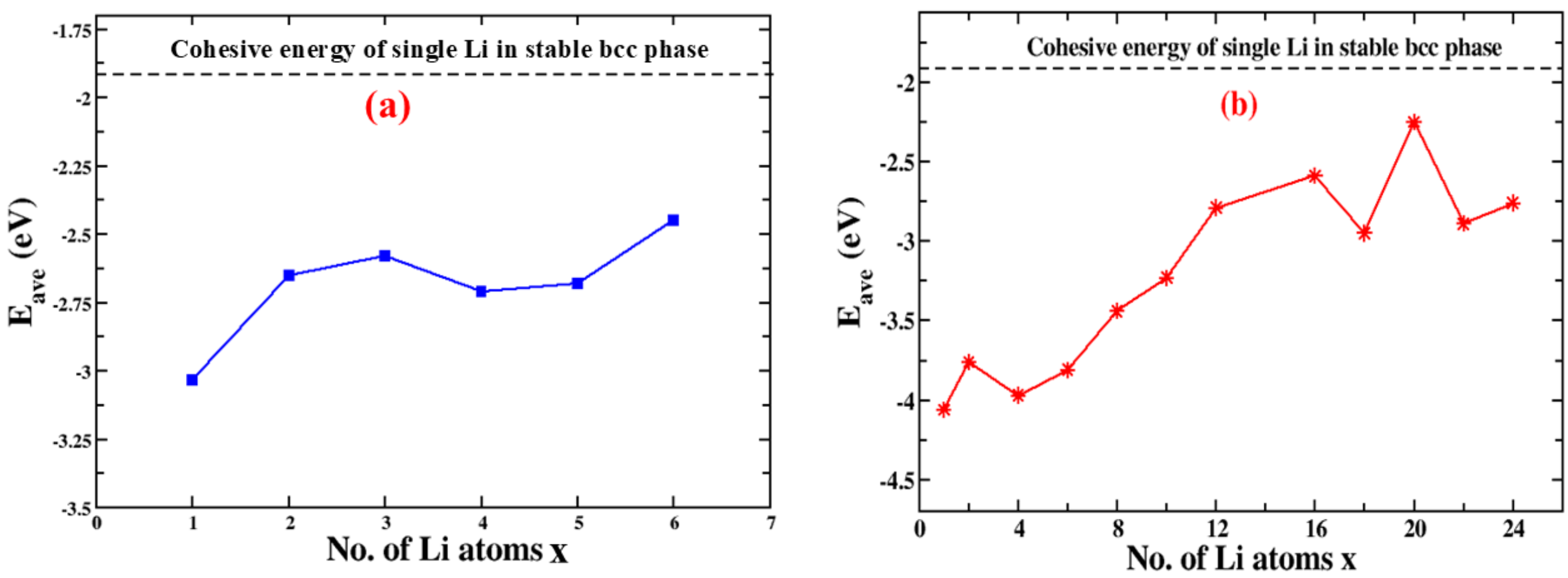

***Fig. 7*** *Average adsorption energy versus different concentration of Li atoms in (a) goldene-I and (b) goldene-II. Here 'x' represents the number of Li atoms adsorbed on the monolayers. Notably,* $x_{max}$ *= 6 for goldene-I (*$3 \times 3 \times 1$ *supercell) and* $x_{max}$ *= 24 for goldene-II (*$2 \times 2 \times 1$ *supercell).*

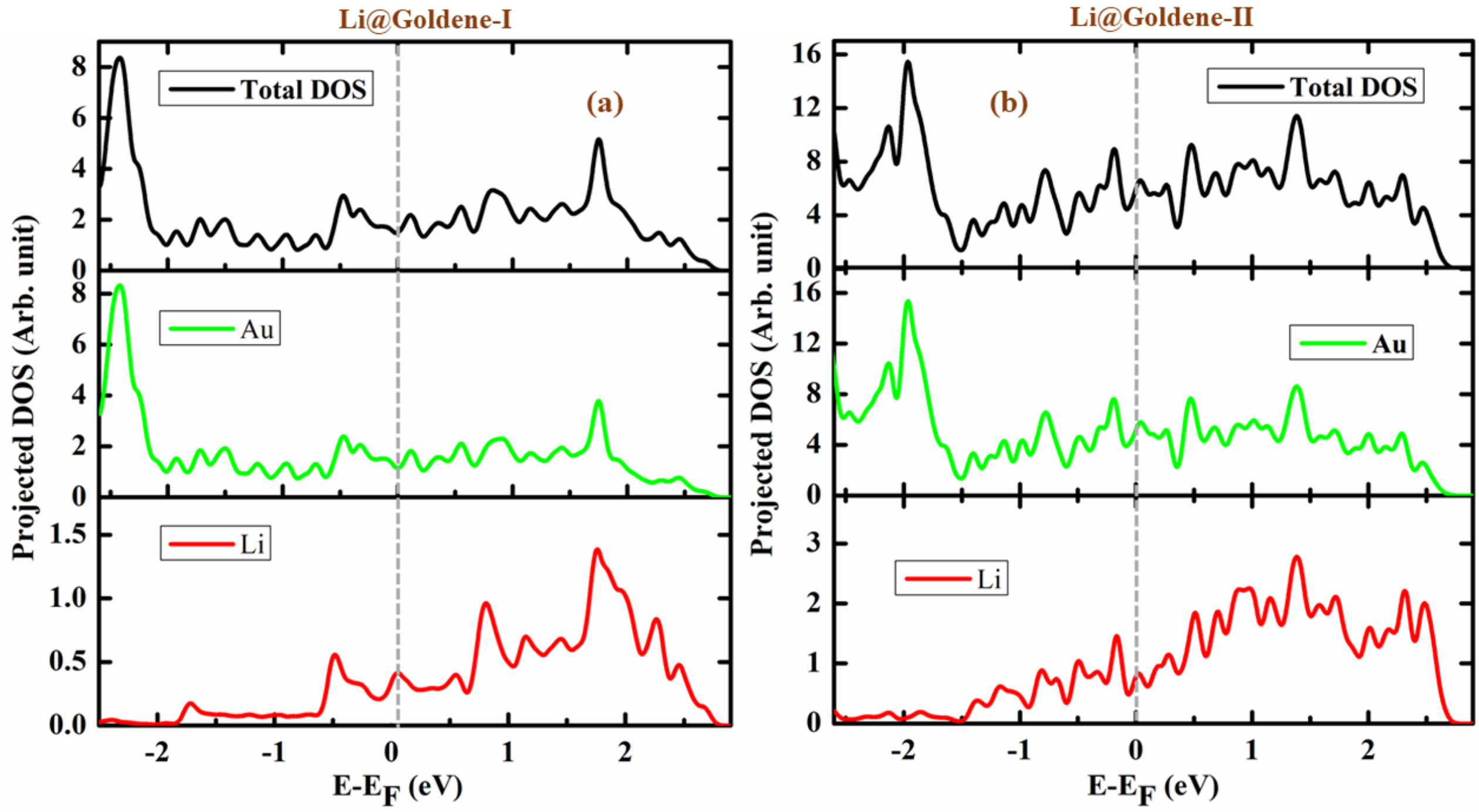

***Fig. 8*** *The projected density of states for maximum Li absorbed on (a) goldene-I and (b) goldene-II monolayer.*

Excellent electronic conductivity is a crucial factor in achieving superior electrochemical performance in electrode materials, particularly during the charging and discharging processes. Both pristine goldene-I and goldene-II monolayers exhibit metallic behaviour, as shown in **Fig. 3**. To explore this further, the electronic properties of these goldene monolayers following Li atom adsorption were analysed by calculating the PDOS for maximum lithium adsorption on both the monolayers, with the separate contributions of Au and Li atoms highlighted in **Fig. 8**. The PDOS plots reveal that a few electronic states cross the Fermi level, confirming the distinct metallic character of both monolayers. For goldene-I, which incorporates 6 Li atoms within a 3×3×1 supercell, the electronic states traverse the Fermi level, as shown in **Fig. 8(a)**, indicating that the monolayer retains its inherent metallic nature. Similarly, in goldene-II, accommodating up to 24 Li atoms in a 2×2×1 supercell, the PDOS presented in **Fig. 8(b),** displays a significant density of states, reflecting an increased number of energy bands crossing the Fermi level. This boost in electronic conductivity results from electron transfer from the Li atoms to the goldene monolayer, enhancing its overall performance. Thus, both monolayers sustain their electronic conductivity even after maximum Li atoms adsorption, making them suitable for high-performance electrode applications.

### 3.3.4. OCV and storage capacity

The average open-circuit voltage (OCV) and storage capacity (C) of rechargeable LIBs are critical parameters for understanding the charge-discharge mechanism. In this study, the following half-cell reaction is considered to illustrate the charge-discharge process of the goldene monolayer:

$$\text{goldene} + x\text{Li}^{1+} + xe^{-} \leftrightarrow \text{Li}_x\text{goldene}\,. \tag{8}$$

The OCV is estimated by calculating the change in Gibbs free energy (ΔG) for the above reaction using the equation:

$$\Delta G = \Delta E + P\Delta V - T\Delta S. \tag{9}$$

In this case, the PΔV term is negligible due to the minimal volume change during lithium adsorption on the goldene surface. Additionally, the entropy contribution (TΔS) is approximately 25 meV at room temperature, which is insignificant compared to the typical adsorption energy (1–2 eV) of Li metals. Thus, the Gibbs free energy change is effectively equal to the internal energy change (ΔE), corresponding to the adsorption energy. Consequently, the average OCV is computed from the average adsorption energy $E_{ave}$. We have calculated OCV profiles using convex-hull method.[60] To determine the OCV profiles of both the goldene monolayers, first we have calculated relative formation energies ($E_{rf}$), and then determined stable intermediate structures for both the phases. To calculate the OCV for different Li atom concentrations, we use the following equation: [61,67]

$$\text{OCV}= [E_{Li_{x_1}@goldene} - E_{Li_{x_2}@goldene} + (x_2 - x_1)E_{Li}]/z\,(x_2 - x_1)e. \tag{10}$$

Where $E_{Li_{x_1}@goldene}$ and $E_{Li_{x_2}@goldene}$ represent the total energy of the systems when there are $x_1$ and $x_2$ number of lithium atoms adsorbed on the supercell of the gold monolayer, respectively. $\mu_{Li}$ is the energy of a single lithium atom in its stable bcc phase, [59] z is the electronic charge of the Li ion. We have plotted relative formation energies with lithium atom concentrations in order to determine the convex hull for optimal configurations among all possible configurations of lithium atoms. The relative formation energies with lithium atom concentrations have been calculated using the following equation: [59,61-62]

$$E_{rf} = E_{Li_n@goldene} - nE_{Li_{1.0}@goldene} - (1-n)E_{goldene.} \tag{11}$$

Where $E_{Li_n@goldene}$ and $E_{Li_{1.0}@goldene}$ represent the total energy of the systems with arbitrary lithium concentration 'n', total energy of the systems with maximum lithium concentration, and the 3rd term in the Eq.(11) represent the total energy of the pristine goldene monolayer, respectively. $\mu_{Li}$ is the total energy of single lithium atom in its stable bcc phase, [59] z is the electronic charge of the Li ion. The maximum number of Li atoms adsorbed on the monolayers corresponds to n = 1 (where x = 6 for goldene-I and x = 24 for goldene-II). As an example, n=0.166 for single lithium atom adsorbed on goldene-I, and n=0.041 for single lithium adsorbed on the goldene-II monolayer. We have determined the stable intermediate structures to plot the convex hull with the help of following criteria: [59]

$$\Delta E = E_{Li_n@goldene} - \frac{(n_h-n)}{(n_h-n_i)} E_{Li_{n_i}@goldene} - \frac{(n-n_i)}{(n_h-n_i)} E_{Li_{n_h}@goldene}. \tag{12}$$

Where n is the arbitrary Li atom concentration, while $n_i$ and $n_h$ are the smallest and the largest concentrations, respectively. $E_{Li_n@goldene}$, $E_{Li_{n_h}@goldene}$, and $E_{Li_{n_i}@goldene}$ are the corresponding total energies of the lithium adsorbed monolayer at an arbitrary concentration, lithium adsorbed monolayer for the smallest concentration of lithium, and lithium adsorbed monolayer for the largest concentration of lithium, respectively. Here 'n' is the smallest when there is only one lithium atom is adsorbed, and n is the largest when maximum number of lithium atoms are adsorbed on both the monolayers. For instance, n=1 is the largest for both the monolayers. Whereas, the smallest value of 'n' is 0.166 for goldene-I and 0.041 for goldene-II. If $\Delta G \leq 0$, then the structure is thermodynamically stable relative to dissociation into other configurations otherwise the structure will break.

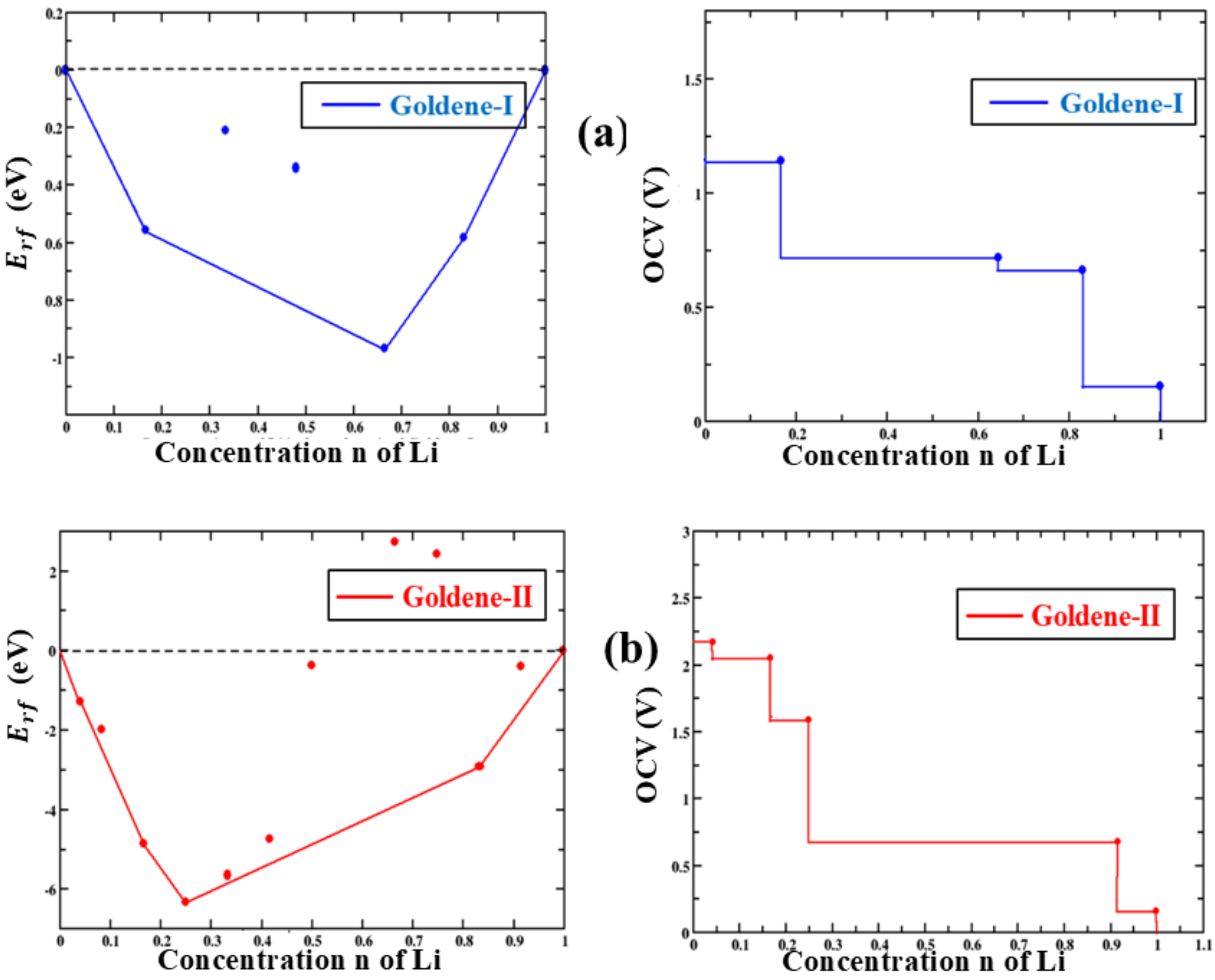

*Figure. 9 Left panel of each of the figure indicates convex hull plot of Relative formation energy as a function of Li and the right panel of each of the figure represents the OCV plot in the formula unit of (a) goldene-I (b) goldene-II monolayer. Notably, here concentraion 'n'=1.0 when maximum number of Li atoms $x_{max}$ =6 and $x_{max}$ = 24 are stored on the goldene-I and goldene-II monolayers, respectively.*

In the left panel of the **Fig. 9**, we have plotted the formation energy as a function of lithium atom concentrations. In the plot dotted points denote relative formation energies for different lithium concentrations/configurations of both the monolayers. A dashed zero energy line is the reference line below which the dotted points represent stable configurations. A discontinuous solid line connecting some most stable configurations is the convex hull. [55, 63] The points on this line represent the configurations are thermodynamically stable. The points above this line but below the reference line are metastable points which are energetically favourable configurations compared to Li in metallic phase but unstable against decomposition into phases on the convex hull. [64, 65] For goldene-I, we see from the left panel of **Fig. 9(a)**, a relatively less deep energy convex hull. The convex hull reaches its minimum to −0.97 eV at n = 0.64 indicating a moderate binding of lithium atoms. For goldene-II phase, we see from the left panel of **Fig. 9(b)**, a V-shape convex hull with a strong-energy minima near n = 0.25 followed by a coexistence of two-phase thermodynamic equilibrium. The deep minima near n = 0.25 determines the electrochemical behaviour of goldene-II. After determining the most stable intermediate structure we have calculated OCV between two successive stable intermediate configurations. From the OCV plot in right panel of **Fig. 9(b)**, we see an initial large voltage around 2.17V followed by an extended plateau shape OCV voltage window. The step-like OCV profile indicates a strong initial lithium insertion. Beyond n = 0.25, an increase in the adsorption energy trend indicates progressively less favourable lithium insertion. We report that our calculated OCV for goldene-I and goldene-II lies in the range of 0.12-1.14 V and 0.79-1.09 V, respectively. Throughout the entire process, the OCV at all concentrations meets the ideal requirements for anode materials, effectively preventing the formation of Li atom dendrites, and minimizing significant damage to equipment during charging and discharging. As a result, these values of OCV enhances both rate capacity and operational safety suggesting that goldene monolayers are well-suited to serve as a promising anode material with a low charge/discharge state.

We also have discussed qualitatively the finite-temperature effect on the relative phase stability of lithium intercalated composite systems at different concentrations of lithium on the convex hull and the step-like OCV profile. In this work, the convex hull was constructed using total energies obtained from density functional theory (DFT) calculations at 0K, which is the standard approach in the first-principles studies of battery materials for determining the phase stability of lithium intercalated composite at different lithium concentrations. However, the practical operating condition of battery is at finite temperature (approximately 300K). At finite temperature, additional free-energy contributions arising from vibrational and configurational entropy may modify the relative stability of Li configurations, particularly at intermediate concentrations. At finite temperature, the OCV can be calculated from the finite temperature Gibbs free energy change for two different lithium concentrations. At finite temperature change in Gibbs free energy can be written as, $\Delta G = \Delta H - T\Delta S$, where $\Delta H$ is the change in enthalpy can be written as, $\Delta H = \Delta H_{DFT} + \Delta H_{vib}$. Where, $\Delta H_{DFT} = \Delta E + P\Delta V$. $\Delta H_{DFT}$ and $\Delta H_{vib}$ are the enthalpy associated with 0K and vibrational enthalpy. As $P\Delta V$ term is negligibly small in solid, $\Delta H_{DFT} \approx \Delta E$ is the change in total energy at finite T, i.e. in our case the adsorption energy that we have calculated using first-principles at 0K which can be written as:

$$\Delta H_{DFT} = E_{DFT}(Li_x goldene) - xE_{DFT}(Li) - E_{DFT}(goldene), \tag{13}$$

where right side of **Eq. (13)** is the adsorption energy that we have calculated using standard first-principles DFT calculations. Now, the total change in Gibbs free energy can be written as:

$$\Delta G(\mathrm{T}) = E_{ads} + \Delta G_{vib} - T\Delta S_{conf}, \tag{14}$$

where 2[nd] term is the sum of contribution of the change in vibrational enthalpy $\Delta H_{vib}$ and the change in vibrational entropy $\Delta S_{vib}$, and the 3[rd] term is the contribution of the change in configurational entropy at finite temperature T. For our system, the vibrational enthalpy ($H_{vib}$), vibrational entropy ($S_{vib}$), vibrational Gibbs free energy $G_{vib}$, and configurational entropy ($S_{conf}$) can be written using a statistical partition function under a given phonon density of states, g( $Li_x goldene, \omega$) as follows:[66,67]

$$H_{vib}(Li_x goldene, T) = \int_0^\infty g(Li_x goldene, \omega) d\omega [\frac{1}{2}\hbar\omega + \frac{\hbar\omega}{(e^{\frac{\hbar\omega}{k_B T}} - 1)}], \tag{15}$$

$$S_{vib}(Li_x goldene, T) = k_B \int_0^\infty g(Li_x goldene, \omega) d\omega [\frac{\hbar\omega / k_B T}{\left(e^{\frac{\hbar\omega}{k_B T}} - 1\right)} - \ln(1 - e^{-(\frac{\hbar\omega}{k_B T})})], \tag{16}$$

$$G_{vib}(Li_x goldene, T) = H_{vib}(Li_x goldene, T) - \mathrm{T}S_{vib}(Li_x goldene, T), \tag{17}$$

$$\Delta S_{conf} = S_{conf}(Li_x goldene) - S_{conf}(goldene) - xS_{conf}(Li), \tag{18}$$

where $\hbar$, $k_B$, $1/\left(e^{\frac{\hbar\omega}{k_B T}} - 1\right)$, $\Omega$ are Planck's constant, Boltzmann constant, and Bose distribution function, and the total number of configurations of possible Li ordering in lithium intercalated composite/system, respectively. The entropy is defined as $S_{conf} = k_B \ln \Omega$. Now, overall $S_{conf}(Li_x goldene)$ is the configurational entropy of the lithium intercalated goldene, $S_{conf}\,(goldene)$ is the configurational entropy of goldene, and $S_{conf}\,(Li)$ is the configurational entropy of Li. The configurational entropy entropy of the lithium intercalated goldene monolayer ($S_{conf}\,(Li_x goldene)$) will only contribute to $\Delta S_{conf}$ because other two terms viz. $S_{conf}\,(goldene)$ *and* $S_{conf}\,(Li)$ are zero as they are most stable phase of pristine goldene and the bcc Li. The vibrational enthalpy originated from the lattice vibrations (phonons), helps gaining a vibrational entropy contribution ($-T\Delta S_{\mathrm{vib}}$), which lowers its free energy. This contribution stabilizes phases low frequency acoustics mode (originates from $\eta = 1/\left(e^{\frac{\hbar\omega}{k_B T}} - 1\right)$, occupation number of Bose-distribution function). Thus, vibrational enthalpy adds a modification in the free energy which is very small compared to $\Delta H_{DFT}$. Earlier Haruyama et al. have reported a very small contribution of an amount of -10 meV to -20 meV of vibrational enthalpy to $\Delta G$(T) for Li-intercalated graphite system. [66] Hence, this small contribution of vibrational enthalpy may make some metastable states near the convex hull thermodynamically stable, may changes the stability ordering, may smoothen the convex hull but does not change the overall thermodynamic trend of stability and step-like OCV profile calculated at 0K. The vibrational entropy originated from the hindered motion of lithium in the intercalated phases as compared to those bulk bcc Li contributes a temperature-dependent stabilization which is dominated by low-frequency modes at 300K that lowers the free energy of softer phases. This small minor correction in free energy changes the stability ordering very small and slightly add a shift in OCV. As temperature increases the configurational entropy favours more disordered phases and haphazard occupation of Li adsorption sites, and solid-solution like behaviour. Configurational entropy also contributes a small amount. For instance, Haruyama et al. had reported that configurational entropy contributes an amount of -20 meV at 300K for lithium-intercalated graphite. [66] So, configurational entropy stabilizes disordered Li arrangements at finite temperature, modifies the stability ordering specifically at intermediate compositions. Overall, the configurational entropy makes convex hull smoother keeping voltage profile almost unchanged with a slight modification in OCV only up to a few tens of meV. Therefore, finite temperature effect would not modify the overall stability ordering too much. [68]

The volumetric storage capacity is determined using the following equation:

$$C = \frac{\mathrm{xnF}}{v_{\mathrm{Li_x goldene}}}, \tag{19}$$

where F is the Faraday constant, n (=1) is the valence state of lithium, and $v_{\mathrm{Li_x goldene}}$ is the effective volume of the Li-adsorbed layered structure. The goldene-I monolayer is capable of adsorbing up to two layers of lithium atoms. However, in the 3×3×1 supercell containing nine gold atoms, only six lithium atoms can be stably accommodated at the T-sites, as illustrated in **Fig. 6(a)**. Further adsorption beyond this limit leads to structural instability and significant distortion of the monolayer, indicating that six Li atoms represent the optimal adsorption capacity for goldene-I.

***Table 2**: Volumetric Capacity of Typical 2D Anode Materials for LIBs.*

| 2D Monolayer Material | Volumetric Capacity (Ah cm$^{-3}$) | Ref. |
|---|---|---|
| Goldene-I (this work) | 0.713 | This work |
| Goldene-II (this work) | 0.783 | This work |
| Graphene | ~0.81 | [69] |
| α- & δ-Borophene | ~1.46-1.70 | [70-72] |
| Phosphorene | 1.40–1.43 | [73] |
| Silicene | 0.95–1.00 | [74] |
| Germanene | ~1.20 | [75] |
| Biphenylene Carbon (BPC) | ~1.05 | [76] |
| $MoS_2$ monolayer | ~1.2 | [77] |
| $Ti_2C$ monolayer | ~1.30 | [78] |

The goldene-II show better storage capacity dues its high lithium-to-gold adsorption ratio, where nearly every Au atom serves as a host for a Li atom. This unique property results in an exceptionally high volumetric storage capacity, making it an ideal candidate for non-portable LIBs, such as those used in grid storage systems. While portable batteries prioritize lightweight designs and high gravimetric capacity, high molecular mass per unit cell of goldene monolayers leads to a modest gravimetric capacity of (88-148) mAh/g, limiting its suitability for portable applications. However, its strength lies in volumetric performance, which is critical for stationary energy storage. For instance, in a 2×2×1 supercell of goldene-II, 24 Au atoms host 24 Li atoms; 8 are positioned on hexagonal motifs (4 on each side) and 16 on trigonal sites (8 on each side), as shown in **Fig. 6(b).** Although its gravimetric capacity may appear low, goldene-II is engineered for volumetric efficiency rather than weight savings. It delivers large energy density in a compact form, making it a game-changer for space-constrained applications, such as grid-scale energy storage. To calculate the volumetric capacity of our systems, we consider the effective volume, which is defined as the product of the in-plane area, and the effective thickness of the Li adsorbed system in its respective supercell. The effective thickness is defined by the following equation:[79]

$$t_{eff} = 2h + 2r_{vdW}. \tag{20}$$

Here, $h$ is the vertical distance between topmost Li atom and Au atom, and $r_{vdW}$ is the average van der Waals radius of Au and Li atoms. The van der Waals radius of Au and Li atoms are $r_{Au}$ =1.66 Å , and $r_{Li}$=1.82 Å, respectively. Thus, $r_{vdW}$ = 1.74 Å. The height between the topmost Li atom and the Au atoms of the matrix monolayer is, $h$ =1.96 Å. The effective thickness of the Li adsorbed goldene-I monolayer is, $t_{eff}$=3.92 + (2 × 1.74)Å = 7.40Å. For goldene-I, the in-plane area is, A=$(\sqrt{3}/2) \times 8.21 \times 8.21$ Å$^2$ = 50.54 Å$^2$. The effective volume of maximally Li adsorbed 3×3×1 goldene-I supercell, $V_{eff} = 50.54 \times 7.40$ Å$^3$ $= 374.05 \times 10^{-24} cm^{-3}$. So total volumetric capacity is calculated to be $C = 0.713$ Ahcm$^{-3}$. Similarly, for goldene-II, the in-plane area, A=$(\sqrt{3}/2) \times 14.21 \times 14.21$ Å$^2$ = 174.86 Å$^2$. The height between the topmost Li atom and the Au atoms of the matrix monolayer is, $h$ = 1.95. The effective thickness of Li adsorbed goldene-II monolayer is, $t_{eff}$=3.90 + (2 × 1.74)Å = 7.80 Å. The effective volume of the Li adsorbed 2×2×1 goldene-II supercell, $V_{eff} = 174.86 \times 7.8$Å$^3$ $= 1363.90 \times 10^{-24} cm^{-3}$. So, the total volumetric capacity is calculated to be $C = 0.783$ Ahcm$^{-3}$. This calculated volumetric capacity has been compared with that of other typical 2D materials for Li-ion batteries in Table 2. Graphite anodes in lithium-ion batteries have a volumetric capacity of ~0.5–0.8 $Ah$/cm$^3$.[81-82] Silicon anodes offer a high theoretical volumetric capacity of approximately 2–3 $Ah/\text{cm}^3$, but they suffer from substantial volume expansion of about 100–200% during full lithiation.[83, 84] In contrast, goldene-II exhibits a significantly lower theoretical volume expansion, limited to around 12–14% at maximum lithiation.

#### 3.3.5. Diffusion barrier

To theoretically evaluate the charge/discharge rate of these goldene phases as potential anode materials, we investigated the diffusion pathways and corresponding energy barriers for a single Li-ion. Since the T-site is the most energetically favourable adsorption site in goldene-I, Li diffusion is expected to occur from one T-site to another. According to structural symmetry, there are two possible migration pathways: PATH-1@goldene-I: Li migrates over the top of Au–Au bonds, as shown in **Fig. 10(a).** PATH-2@goldene-I: Li moves over the centre of Au atoms, as depicted in **Fig. 10(b).** The diffusion barrier for goldene-I for PATH-1 is only 15 meV, and for PATH-2 is 180 meV, which is much smaller than the reported values in other 2D monolayers, as reported in **Table 4**.

The relatively tiny diffusion barrier along the Path-1 compared to the Path-2 for goldene-I can be understood from adsorption energy and bonding characteristics at different adsorption sites. We have shown in **Fig. 4**, the adsorption energy and the corresponding histogram. We see the adsorption energy at T-site ( -3.036 eV) and B-site (-3.021 eV) are almost same. T-site is the most stable adsorption site. Whereas, Au-site is least stable site. Also, in the **Table 3**, we have shown the ICOHP values of Li-Au bonds for all the symmetric sites. The more negative the ICOHP, the stronger the bonding characteristics. From the **Table 3**, it is clear that significantly large negative value of -1.162 eV at the T-site binds lithium atom strongly. The ICOHP values are in consistent with adsorption energy values. For PATH-1 (T→B→T), the migration barrier peaks at B-site. Thus, for PATH-1, the barrier height will be approximately the energy difference between T-site and B-site configurations, which is the same as adsorption energy difference between T-site and B-site, i.e., 15 meV (see **Table 1**). Also, from the EDD plot in the **Fig. 5**, we see that at the T-site and B-site there are delocalisation of electrons after lithium adsorption on the matrix. This delocalisation of charge also offers a smooth migration channel along PATH-1 (T→B→T). So, the diffusion barrier we have got is 15 meV which is ultra-low. But, along the Path-2 (T→Au→T) lithium migrates from one T-site to nearest T-site via Au-site. The Au-site is the least stable adsorption site, exhibiting a higher adsorption energy of 180 meV relative to the T-site (see **Table 1**). So, due to the significant adsorption energy difference between the T-site and Au-site, lithium ions face a diffusion barrier of 180 meV along the Path-2 for goldene-I (see **Fig. 10 (d)**).

*Table 3: ICOHP values for Li-Au bonds for different adsorption sites determining bonding strength of Li-Au in goldene-I and goldene-II.*

| | Goldene-I | | | Goldene-II | | |
|---|---|---|---|---|---|---|
| ICOHP (in eV) | T-site | B-site | Au-site | H-site | B-site | T-site |
| | -1.162 | -0.543 | -0.529 | -1.584 | -1.319 | -1.169 |

For goldene-II, the H-site is a more favourable site as can be seen from ICOHP values (see **Table 3**) and adsorption energy (see **Table 1**). There is only one pathway from one H-site to another H-site, named PATH-1@goldene-II. The calculated diffusion barrier height is 0.59 eV for goldene-II when Li ions migrate from one H-site to another. This NEB barrier height (for path H→ B→H) is in consistent with the adsorption energy difference between H-site and B-site. In **Table 4**, we have provided the diffusion barrier of different 2D materials. The diffusion coefficient of a metal atom plays a critical role in determining the charge/discharge rate when evaluating a material as a negative electrode for rechargeable batteries. Although goldene-II shows a comparatively higher diffusion barrier due to this strong binding, the barrier is not excessively high. In fact, its diffusion barrier remains within the practical range for anode materials-comparable to or even lower than widely used systems, such as graphite and graphene, which exhibit diffusion barriers of around 0.3-0.8 eV.[59] On the other hand, goldene-I exhibits very low diffusion barriers due to weaker Li adsorption, which indeed facilitates ultrafast ion migration. However, this weak binding sometime also leads to poor Li stability, limited charge transfer, and lower storage capacity.

We have checked the convergence test of this ultra-low diffusion barrier of 15 meV for goldene-I with different convergence parameters such as plane wave cutoff energy, k-point sampling, and number of images used for CI-NEB method. The detailed results of the convergence test of diffusion barrier are provided in **Table S1**, **Table S2**, and **Table S3 (**see in **SI**). This ultra-low (15 meV) diffusion barrier of goldene-I from different convergence test demonstrates the reliability of our obtained diffusion barrier which shows a little variation (approximately 13-18 meV) around 15 meV. An important point to note that, despite of strong adsorption energy (-3 eV), goldene-I shows almost barrierless for lithium migration. The diffusion barrier solely depends on the potential energy surface. If the potential energy surface is smooth along the migration pathways the diffusion barrier becomes low though strength of adsorption be large. Similar ultra-low diffusion barrier for lithium migration were predicted in other two-dimensional materials such as $Ti_2N$ (17 meV) [102], $Sc_2C$ (18meV) [101], $Ca_2B_2$ (16 meV) [85], $Ti_2B_2$ (17 meV) [86], $Zr_2B_2$ (17 meV) [87], etc. This ultra-low diffusion barrier even smaller than room temperature thermal energy (25 meV) indicates almost barrierless migration of lithium ion, and rapid diffusion kinetics.

***Table 4****: Diffusion barrier of a few 2D anode materials for LIBs.*

| 2D materials | diffusion barrier (meV) | References |
|---|---|---|
| Graphene & its derivatives | 320-850 | [88] |
| Borophene | 600-820 | [38, 69] |
| Phosphorene | 130-760 | [89] |
| Biphenylene | 230 | [90] |
| BCN- biphenylene | 650 | [91] |
| $Be_2C$ | 110 | [92] |
| $BeP_2$ | 150 | [93] |
| AsP | 170 | [94] |
| FeAs | 380 | [95] |
| FeSe | 160 | [96] |
| $VS_2$ | 220 | [97] |
| SiC | 570 | [98] |
| BGe | 200 | [99] |
| $NiC_3$ | 510 | [100] |
| $Sc_2C$ | 18 | [101] |
| $Ti_2N$ | 17 | [102] |
| $Mn_2C$ | 24 | [103] |
| $Ca_2C$ | 27 | [104] |
| GaN | 79 | [105] |
| TaB | 90 | [106] |
| goldene-I | 15 | This work |

| goldene-II | 590 | This work |
|---|---|---|

The temperature-dependent molecular jump rate has been calculated using the Arrhenius equation, with the diffusion coefficient (D) expressed as:

$$D = \frac{1}{4} \Gamma \times l^2, \quad (21)$$

where $l$ represents the nearest hopping distance, and Γ denotes the jump rate, defined by:

$$\Gamma = f_0 e^{-E_b/k_B T}. \quad (22)$$

Here, $f_0$ is the hopping frequency of Li atom within the monolayer, $k_B$ is Boltzmann constant, $T$ is the absolute temperature (300 K), and $E_b$ is the diffusion barrier. Here, the $f_0$ value for Li atom on 2D materials mostly lies in the range of $10^{12}$-$10^{14}$ $s^{-1}$. [107-108] For goldene-I, the energy barrier is low at 15 meV, compared to $k_B T$ at 300 K, which is approximately 25 meV. The exponential term is close to 1 in the Arrhenius **eq. (22),** meaning the diffusion rate is proportional to hopping frequency, which makes physical sense for fast diffusion $5.6\times10^{13}$ $s^{-1}$. Using the 2D diffusion **eq. (22),** while considering the nearest hopping distance of 1.47 Å, the diffusion coefficient at 300 K is reported as $3.025\times10^{-3}$ cm²/s, reflecting its high ionic mobility. In contrast, for goldene-II, the energy barrier is higher at 0.59 eV, significantly exceeding $k_B T$ (0.025 eV at 300 K). Here, the exponential term is far from 1, indicating that the hopping frequency is a smaller fraction of the attempt frequency. This aligns with moderate diffusion behaviour. The diffusion coefficient is found to be $4.84\times10^{-11}$cm²/s at 300 K.

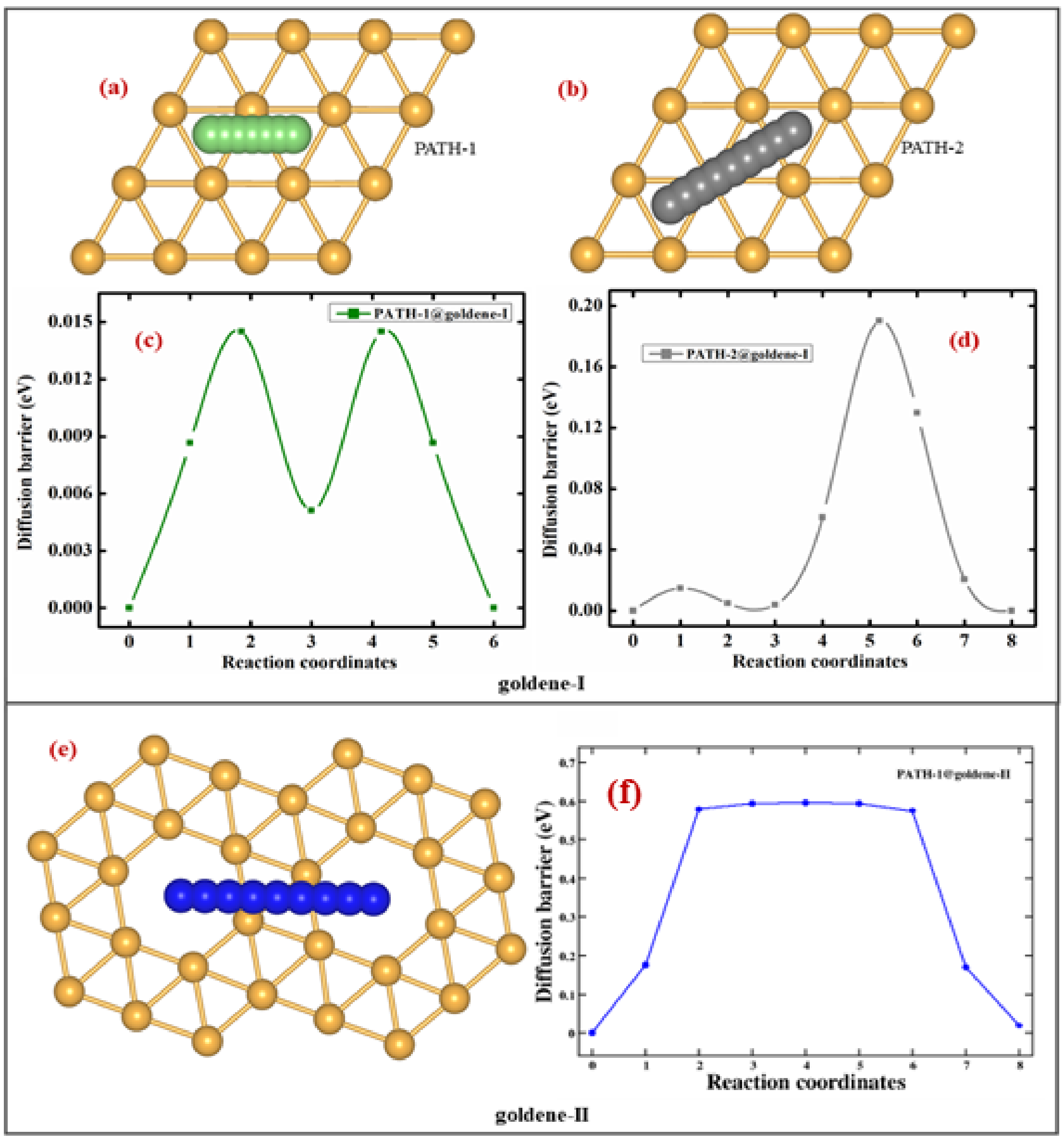

*Fig. 10 Diffusion paths and their corresponding energy barrier; (a) and (b) are reaction paths for two different Path for goldene-I and (b) and (d) diffusion barrier for goldene-I, and (e) and (f) trajectory and energy barrier for goldene-II respectively.*

### 3.3.6. AIMD of maximum Li adsorbed goldene monolayers

To evaluate the thermal stability and sustainability of lithium adsorbed goldene monolayers, we performed AIMD simulations at 300 K, corresponding to maximum lithium adsorption. These simulations spanned over 5,000 fs with a time step of 1 fs, allowing for a detailed observation of the dynamic evolution of the system. The free energy fluctuations throughout this duration have been monitored and plotted, providing a quantitative indicator of structural stability. Geometric snapshots after 5000 fs, presented in **Fig. 10**, demonstrate that both goldene-I and goldene-II retain their structural integrity under maximum Li adsorption. In goldene-I, the Li atoms exhibit minimal displacement from their initial adsorption sites, reflecting a robust interaction with the goldene lattices can be seen in **Fig. 11(a)**. Notably, no significant structural distortion or lattice collapse is observed in either monolayer, underscoring their resilience at ambient conditions.

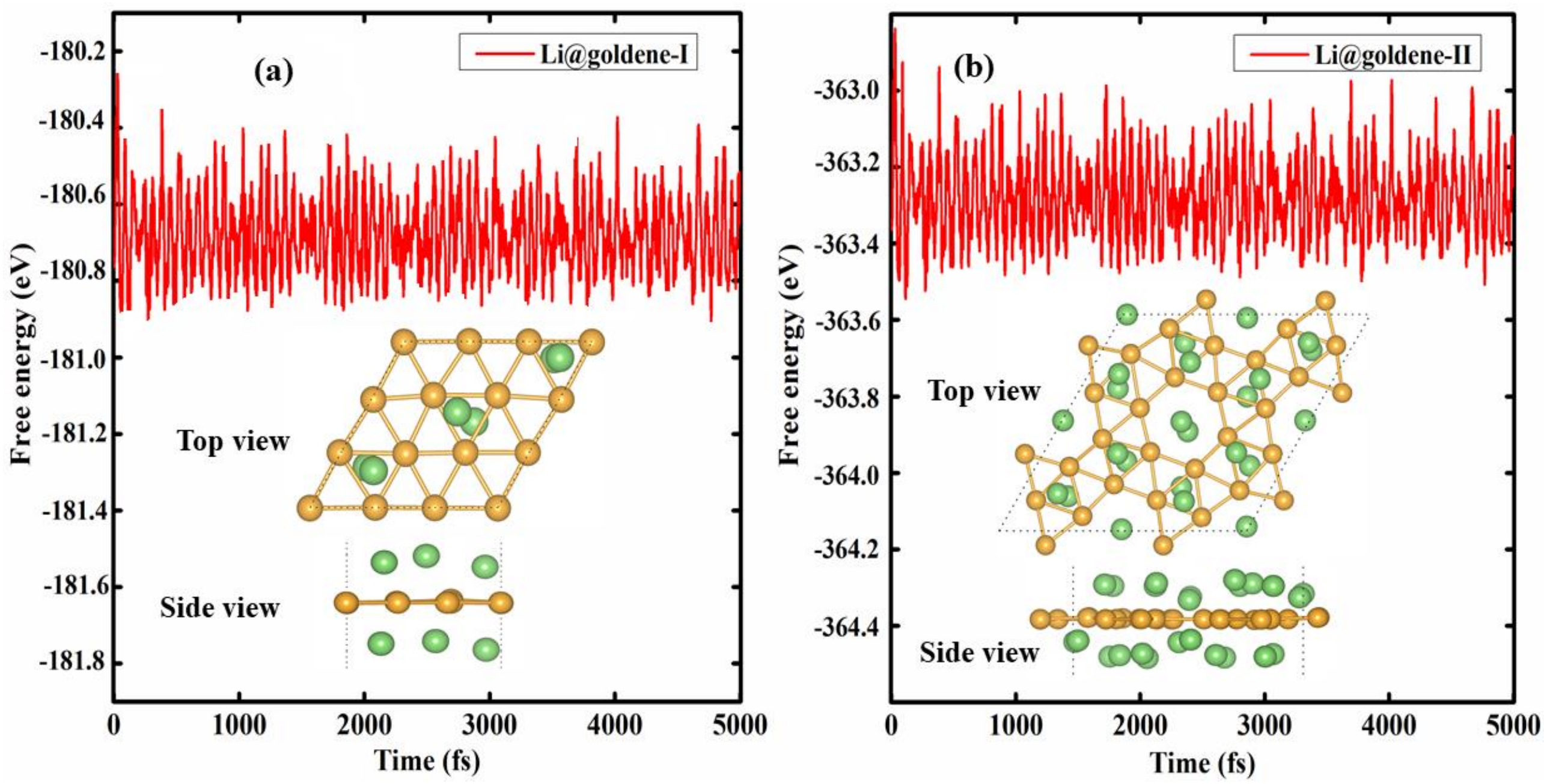

***Fig. 11*** *Free energy as a function of time step from AIMD simulations conducted at 300 K, along with top and side view snapshots of the optimized geometries for maximally Li adsorbed (a) goldene-I and (b) goldene-II after 5000 fs.*

On the other hand, in goldene-II, Li has a displacement at the H-site; yet, this shift does not induce lattice distortion, as can be seen in **Fig. 11(b)**. A similar minor deviation is observed at the T-site, but it remains insufficient to compromise the overall framework. The absence of pronounced deformations indicates that both goldene monolayers can sustainably accommodate Li adsorption without undergoing substantial structural changes, even under dynamic conditions at 300 K. This stability is likely attributable to the intrinsic properties of the goldene lattice, which effectively balances flexibility and rigidity to support Li binding.

## Conclusion

Drawing from the recent experimental synthesis and exfoliation of goldene, a gold monolayer derived from the nano-laminated ternary ceramic phase $Ti_3AuC_2$, we present a theoretical investigation of the reported goldene-I phase alongside a newly proposed goldene-II structure. The goldene-II, characterized by a lattice of triangular gold atom motifs, exhibits exceptional structural stability and distinct geometric properties within a two-dimensional framework. In contrast, the proposed goldene-II phase, incorporating a hybrid lattice of triangular and hexagonal motifs with periodic pores, demonstrates robust structural integrity and mechanical stability, even under lithium adsorption, as validated through first-principles calculations. Both phases exhibit metallic behavior, as confirmed by their electronic band structures and projected density of states, which reveal dispersive bands crossing the Fermi level. Our computational analysis of electrochemical properties indicates that Goldene-I achieves a volumetric capacity of 0.713 Ah/cm³, while goldene-II reaches 0.783 Ah/cm³, confirming the high suitability of gold monolayer for lithium storage as alternative anode material in LIBs. Furthermore, diffusion barrier calculations yield an exceptionally low energy barrier of 15 meV for goldene-I. Thus, despite having slightly higher diffusion barriers, goldene-II provides a better overall balance between strong Li adsorption (stability), high storage capacity, and acceptable ion mobility, making it an alternative anode material for LIBs.

### CRediT authorship contribution statement

**Ajay Kumar**: Investigation, Software, Writing – original draft and analysis of results.
**Pritam Samanta:** Investigation, Software, Writing – original draft and analysis of results.
**Prakash Parida**: Validation and Supervision.

### Declaration of competing interest

The authors declare no conflict of interest.

### Acknowledgements

AK thanks University Grants Commission (UGC), New Delhi, Government of India, for financial support in the form of a Senior Research Fellowship (DEC18-512569-ACTIVE). PP thanks DST-SERB, Government of India, for ECRA project (ECR/2017/003305).

### Data availability

Data will be made available on request.

# Supplementary information (SI)

## Electrochemical Performance of Gold Monolayers for Lithium-Ion Batteries: A First-Principles Study

Ajay Kumar[a], Pritam Samanta[b] and Prakash Parida[b]*
[a]Department of Physics, Government Degree College, Chatroo, Kishtwar, 182205, Jammu and Kashmir, India
[b]Department of Physics, Indian Institute of Technology Patna, Bihta, 801106, Bihar, India
*Corresponding authors: pparida@iitp.ac.in

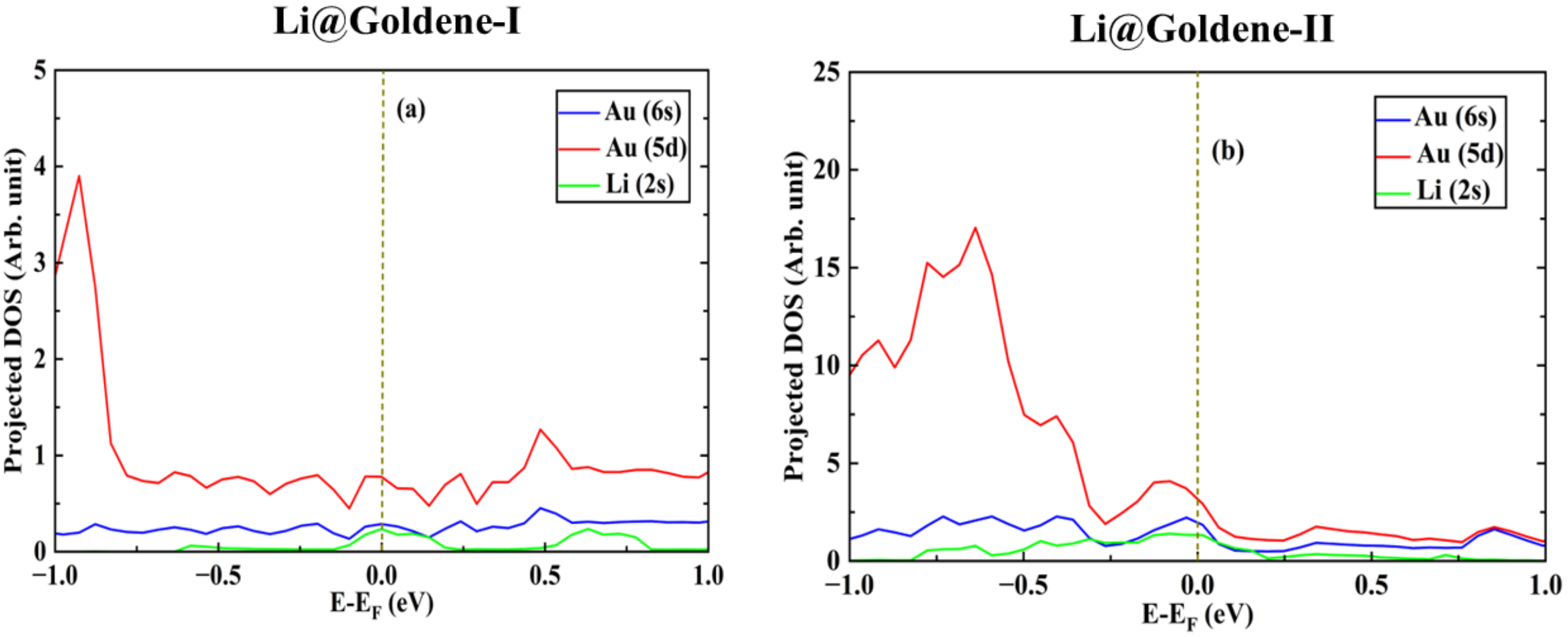

***Fig. S1:*** *The projected density of states for single Li absorbed on (a) goldene-I and (b) goldene-II monolayer at the corresponding most stable adsorption site (T-site for goldene-I and H-site for goldene-II).*

| Cutoff energy (eV) | Diffusion barrier (meV) |
|---|---|
| 400 | 15 |
| 450 | 15 |
| 500 | 17 |
| 550 | 16 |
| 600 | 15 |
| 650 | 14 |
| 700 | 15 |
| 750 | 14 |
| 800 | 14 |
| 850 | 15 |
| 900 | 14 |
| 950 | 16 |
| 1000 | 15 |

***Table S1:*** *Convergence test table of diffusion barrier with respect to different cutoff energies in the range of 400-1000 eV.*

***Table S2:*** *Convergence test table of diffusion barrier with respect to different k-point sampling in the range 4×4×1-15×15×1.*

| k-point grid | Diffusion barrier (eV) |
|---|---|
| 4×4×1 | 15 |
| 5×5×1 | 17 |
| 6×6×1 | 13 |
| 7×7×1 | 15 |
| 8×8×1 | 16 |
| 9×9×1 | 15 |
| 10×10×1 | 16 |
| 11×11×1 | 15 |
| 13×13×1 | 15 |
| 15×15×1 | 13 |

***Table S3:*** *Convergence test table of diffusion barrier with respect to number of intermediate images in the range of 3-9 images.*

| No. of Images | Diffusion barrier (meV) |
|---|---|
| 3 | 16 |
| 5 | 15 |
| 7 | 18 |
| 9 | 14 |